\documentclass{article}
\usepackage[utf8]{inputenc}
\usepackage{dcolumn}
\usepackage{graphics}
\usepackage{enumerate}
\usepackage{graphicx}
\usepackage{subcaption}

\usepackage{amssymb}
\usepackage{amsmath}
\usepackage{siunitx}
\usepackage{bm}
\usepackage{multirow}
\usepackage{url}
\usepackage{color}
\usepackage{authblk}

\title{Explainable and Optimally Configured Artificial Neural Networks for Attack Detection in Smart Homes}

\author[1, 4]{Shaleeza Sohail}
\author[2]{Zongwen Fan}
\author[1]{Xin Gu}
\author[3]{Fariza Sabrina}
\affil[1]{The University of Newcastle, Australia}
\affil[2]{Huaqiao University, China}
\affil[3]{Central Queensland University, Australia}
\affil[4]{Kings Own Institute, Australia}
\date{May 2022}

\begin{document}
\maketitle
\begin{abstract}
In recent years cybersecurity has become a major concern in adaptation of smart applications. Specially, in smart homes where a large number of IoT devices are used having a secure and trusted mechanisms can provide peace of mind for users. Accurate detection of cyber attacks is crucial, however precise identification of the type of attacks plays a huge role if devising the countermeasure for protecting the system. Artificial Neural Networks (ANN) have provided promising results for detecting any security attacks for smart applications. However, due to complex nature of the model used for this technique it is not easy for normal users to trust ANN based security solutions. Also, selection of right hyperparameters for ANN architecture plays a crucial role in the accurate detection of security attacks, especially when it come to identifying the subcategories of attacks. In this paper, we propose a model that considers both the issues of explainability of ANN model and the hyperparameter selection for this approach to be easily trusted and adapted by users of smart home applications. Also, our approach considers a subset of the dataset for optimal selection of hyperparamters to reduce the overhead of the process of ANN architecture design. Distinctively this paper focuses on configuration, performance and evaluation of ANN architecture for identification of five categorical attacks and nine subcategorical attacks. Using a very recent IoT dataset our approach showed high performance for intrusion detection with 99.9\%, 99.7\%, and 97.7\% accuracy for Binary, Category, and Subcategory level classification of attacks. 
\end{abstract}

\section{Introduction}
\label{intro}
The Internet of things (IoT) has been recognised as a disruptive technology with a huge impact in our everyday life. Some widely used applications are smart home, smart health, smart city, smart grid, smart agriculture, smart cars, logistic tracking to name a few \cite{lee2020}. IoT devices are connected via the Internet that collect and exchange data with each other for a particular purpose \cite{yang2017survey}. Due to a huge demand for IoT based applications, the IoT devices are continuously evolving in terms of low cost, easy handling, small form factor and sensor precision. However, security and privacy has not been given a priority while designing these devices \cite{thamilarasu2019deepLearning}. Therefore a huge amount of research has focused on securing IoT based applications; however, a lot more needs to be done to protect IoT networks from intruders.

\begin{figure}
\centering
\includegraphics[width=200pt]{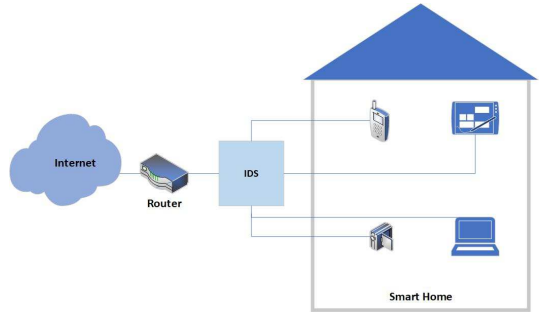}
\caption{Smart home scenario}\label{fig:MAMID}
\end{figure}

Smart home is one of the applications that has played an important role in massive growth of IoT usage. Smart home allows individuals to fully connect their home appliances/devices with each other and also enable the owner to control the devices remotely, as shown in Figure \ref{fig:MAMID}. That provides unprecedented control and comfort. But this benefit comes with a huge security risk, if adequate security measures are not taken into account. In a smart home environment, one of the biggest challenges is in terms of users inability to understand, appreciate and take security precautions. Most IoT devices used in smart homes are designed with an extended set of sensors and actuators. For example, smart light may come with a microphone that may leave the IoT system vulnerable to security and privacy intrusions. A common household resident may not have knowledge and skills to interpret and safeguard the IoT system \cite{dolan2020Proactively}. Hence, there is a huge need for an automated mechanism to protect smart homes from security and privacy intrusions. 

In recent years, many research efforts are focused on automated intrusion detection systems using machine learning. The essence of a machine learning based system is to train the activity model based on the signature, pattern or characteristics of attacks which will be compared against the new traffic behaviour. A number of different supervised and unsupervised machine learning algorithms such as random forest, decision tree, support vector machine, long short-term memory (a type of neuron network), Artificial Neural Networks (ANN) could be used for this purpose. ANN is one of the machine learning algorithms that has been widely used for classification tasks across many domains, due to its black box nature of adapting to the underlying system \cite{zhang2000neuralnetwork}. This property makes ANN highly suitable for network traffic intrusion detection when a high dimensional data set is involved due to its robust ability of model building when the pattern is buried \cite{dreiseitl2002LogisticRA}. A very important aspect of ANN's high performance lies in optimal selection of its hyperparameters for that dataset \cite{choras20217ID}. Considerable effort is required for finding the optimal setup, which is a non-trivial task as it not only affects performance but can be a very time consuming exercise due to availability of a large number of options for each parameter involved in the setup.  

In this paper we propose a novel multi-layer ANN based intrusion detection approach for smart homes. This approach is capable of identifying network attacks at sub-categorical level, which has not been explored extensively in the existing literature. The strength of the approach is its dynamic ability to fine tune ANN hyperparameters to detect attacks with high precision. We would like to emphasise that this selection of hyper parameters is based on detection accuracy, computational overhead and time constraint. The accuracy of prediction is important as missing any attack in case of wrong detection can be very damaging. Computation overhead plays an important role as finding the most optimal solution using extensive computational resources may not be feasible and affordable. Time is one of the constraints as such detection has to be real time and any model that takes a long time to train or detect is not suitable for intrusion detection in this scenario. We also evaluate the selection of the hyperparameters to analyse efficiency of the proposed model. An important thing to mention is that we evaluate our approach on a new IoT dataset, IoTID20 \cite{Ullah2020} as the dataset shows recent network traffic characteristics (collected in 2020). 

The main contribution of this paper is:
\begin{itemize}
    \item We are providing an optimised ANN model with low overhead.
    \item Our approach shows high performance for security attack detection at Subcategory level for a recently captured IoT dataset which shows applicability of the approach for detection of new attacks.
    \item We analyse the explainability of the ANN model which can be helpful for user understanding and trust building.
\end{itemize}

The paper is organised as follows: section \ref{relatedwork} discusses related literature in ANN based intrusion detection system and also the research carried out using IOTID20 dataset. Section \ref{MAMID} provides complete details of our proposed multi tiered ANN model for intrusion detection. The hyper parameter selection is given in section \ref{hyperparameterANN}. Section \ref{results} provides the categorisation results and analysis for intrusion detection purpose. We conclude our paper in section \ref{conclusion} and identify future work options.  

\section{Related Work} 
\label{relatedwork}
There has been a huge amount of research actively going on in the area of Cyber Security where machine learning algorithms are used for anomaly and attack detection. We provide a brief review of some of the literature that is relevant to our work in the following subsections. 

\subsection{ANN Based Intrusion Detection} \label{ANNbasedID}
As mentioned above, machine learning based anomaly and intrusion detection is an active research area; some of the commonly used algorithms are Random Forest (RF),K-Nearest Neighbors (KNN), Support Vector Machine (SVM), Multi-Layered Perceptions (MLP), Artificial neural network (ANN) etc. \cite{islam2020deep}. Among all these different machine learning algorithms, ANN has been considered as one of the most efficient algorithms for anomaly or intrusion detection \cite{tsaih2018ann} \cite{choras20217ID}.

Taher et al. \cite{taher2019NID} compared the performance of SVM and ANN in detection network intrusion and found the ANN performs better than SVM. Authors used NSL-KDD data-set in this work.

Choras et al. \cite{choras20217ID} proposed an ANN based system for intrusion detection for multi-class classification. In this work, the authors tried various possible options of different hyperparameters to come up with the most efficient system. The authors argue that selection of hyperparameters plays a very important role in classification results and a little change in hyperparameter setup has a major impact on accuracy. The hyperparameters that were fine tuned in this work were activation function, the optimiser, number of epochs, number of neurons and batch size. The performance of the proposed model was evaluated using two popular datasets (NSL-KDD and CICIDS2017). Experimental results show that the most effective model achieves an accuracy of 99.909\% for multi-class attack detection.

Hodo et al. \cite{hodo2016ann} presented an ANN based network intrusion detection system for IoT networks. This work focused on binary classification - detecting normal and attack traffic (DDoS/DoS). The performance of the proposed system was tested in a simulated IoT network and the result shows 99.4\% accuracy in detecting Distributed Denial of Service (DDoS) and Denial of Service (DoS) attacks.

Subba et al. \cite{subba2016ann} proposed an ANN based intrusion detection system that could be easily deployed for real-time intrusion detection. The authors argued that some of the existing machine learning based systems may achieve high performance in intrusion detection, but suffer from high complexity and computational overhead. So in this work, authors proposed a three layer ANN model (with one hidden layer) to minimize the computational overhead for training and execution of the system. The performance of the model has been analysed using NSL-KDD dataset. The results from performance evaluation show that this proposed system performs better than Naive Bayes based model and similar as SVM and C4.5 based IDS models.

From the above discussion, it is apparent that ANN is a good choice for intrusion detection purpose but faces a challenge of being complex for optimal selection of hyperparameters. 

\subsection{IoT Dataset} \label{IoTDataset}
When machine learning algorithms are used for attack detection then datasets become very crucial as training and testing of these algorithms cannot be carried out without a dataset. Recently, a number of datasets have been used by researchers for this purpose, like UNSW-NB15 \cite{unsw15} and Bot-IoT \cite{botiot}. We have chosen the IoTID20 dataset for training and testing our ANN model as this is one of the newest dataset collected in the IoT environment and it may depict more realistic and up to date network traffic characteristics. Another important aspect to mention is that as this dataset was collected and made available for use very recently, there are a handful of research utilising this dataset for experimentation. Moreover, none of the existing research (until June 2021) was able to achieve Subcategory classification accuracy of more than 90\% for this dataset.  We have considered this dataset to analyse our IoT intrusion detection model as there is a scope of improvement. 

Ullah and Mahmoud \cite{Ullah2020} collected and experimented with this dataset while considering a number of machine learning approaches. Anomalous activities are detected at binary, category and subcategory level. A number of machine learning algorithms (SVM, Gaussian NB, LDA, Logistic Regression, Decision Tree, and Random Forest) and an ensemble method are used for IoT attack detection with varying results. For binary and category classification Decision Tree, Random Forest and Ensemble methods produced very good results with F-Score of more than 95\% in most cases. However, when it comes to subcategory classification none of the algorithms was able to get F-Score and Accuracy more than 88\%. In security scenarios, getting high accuracy for attack detection is highly recommended and the intrusion detection system with 88\% accuracy may not be relied upon completely. Hence, it can be deduced that there is a room for improvement to the machine learning model in order to achieve high accuracy and precision for subcategory based attack detection using this dataset.    

The importance of feature selection has been analysed for anomaly based intrusion detection approaches and the IoTID20 dataset is used for testing the hybrid feature selection approach using Random Forest algorithm \cite{Maniriho2020}. The authors only discussed partial results for three attack categories indicating more than 99.9\% accuracy. The accuracy for detection of some attack categories is very high which shows the strength of the hybrid feature selection approach that only uses a subset of features which are common among the features selected using Information Gain and Gain Ration methods. However, the applicability of such feature selection approach cannot be justified unless complete results are provided and analysed for all category and subcategory based attacks.  

Qaddoura et al. \cite{qaddoura2021} have proposed a multistage approach for anomaly detection for imbalanced datasets like IoTID20. The authors used approaches like oversampling and K-means++ clustering  in order to mitigate the impact of imbalance among the examples for different categories. The results showed good accuracy for binary classification only. The classification of categories and sub categories was not undertaken by this research work and hence, the effectiveness of this approach for multi-class classification in the IoTID20 dataset cannot be assessed. Similarly, Anjum \cite{farah2020} has discussed cross dataset evaluation considering IoTID20 and Bot-IoT datasets. However, all the attack detection was only for binary or category level and subcategory level classification was not considered by the author. 

\section{Multi-tiered ANN Model for Intrusion Detection (MAMID)} 
\label{MAMID}

\begin{figure}
\centering
\includegraphics[width=250pt]{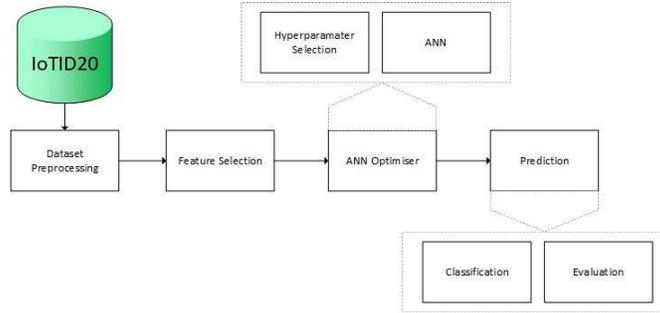}
\caption{Proposed multi tiered ANN model for intrusion detection (MAMID)}\label{figure-MAMID}
\end{figure}

In this section we describe our system that uses ANN for intrusion detection by selecting hyperparameters for high prediction accuracy, low computational overhead and real time detection, the proposed model is shown in Figure \ref{figure-MAMID}. The dataset is preprocessed in order to eliminate any redundant data and missing values. After the preprocessing, feature selection is performed in order to find the minimum set of features that can be effectively used for classification. The data is then passed to the ANN optimiser that finds the optimal neural network topology and hyperparameters for configuring the neural network for training and testing. Not only the prediction accuracy but also the computational overhead and time complexity is considered in order to apply this approach to smart home scenario that requires real time detection and may not have access to high end servers. The following subsections discuss the main components of our system in detail.

\subsection{ANN}
\begin{figure*}[t]
\centering
\includegraphics[width=0.5\textwidth]{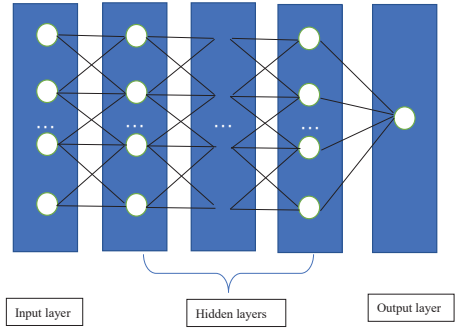} %
\caption{A structure of a classic artificial neural network, including one input layer, multiple hidden layers, and one output layer}
\label{fig:ann}
\end{figure*}

The ANN model consists of three types of layers \cite{mehlig2019artificial}, input layer, hidden layer and output layer, as shown in Fig. \ref{fig:ann}. In each layer, there are many nodes called neurons. The input layer contains the input data feeding into the model. The number of neurons is the same as the number of input features. Each neuron from the input layer interconnects with one or more neurons in the second layer with a weight. The output of the second layer will go to the next layer until the output layer.

Given a three-layer ANN, the weights and a bias term between the input layer and the hidden layer are $\bm{W}^1$($x_1, x_2, x_3$) and $b^1$, respectively. The inner product between the weight matrix and the input data is transformed by an activation function $g(x)$ (e.g. the Sigmoid activation function in Eq. \ref{eq:sigmoid}). The mapped values are fed into the next layer until the output layer. For binary classification, there is only one neurons in the output layer. The feed-forwarded process can be formulated as given in Eq.~(\ref{eq:ann}).
\begin{equation}\label{eq:ann}
f(X) = \bm{W}^2g(\bm{W}^1\bm{X}+b^1) + b^2,
\end{equation}
\begin{equation}\label{eq:sigmoid}
g(x) =  1 / (1+exp(-x)),
\end{equation}
where $g(\cdot)$ is a Sigmoid activation function, it can also be Softmax, Tanh, ReLU activation functions etc, $\bm{W}^1$ and $\bm{W}^2$ are weight matrices for the input and hidden layers, respectively, and $b^1$ and $b^2$ are bias terms added to the hidden and output layers, respectively.

\subsection{ANN Optimiser}
\begin{figure}
\centering
\includegraphics[width=250pt]{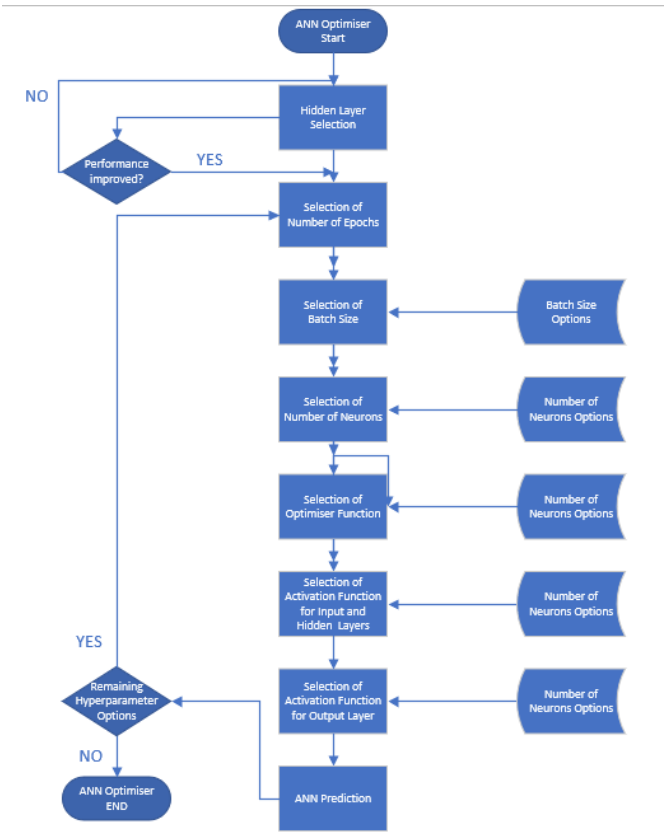}
\caption{ANN Optimiser}\label{figure-fchart}
\end{figure}

The optimisation of ANN setup is the most important aspect when it comes to the accuracy of the prediction and the computational overhead involved in the process \cite{choras20217ID}. With the correct selection of topology and hyperparameters, the model can achieve the highest accuracy of the ANN classifiers. Figure \ref{figure-fchart} shows hyperparameter selection process by ANN optimiser. A general description of the hyperparameters is given below. The actual options for each hyperparameter that we experimented with are given in Section \ref{hyperparameterANN}. 

Optimisation methods are crucial when it comes to the effective training of the neural networks. Optimisation methods are used to reduce losses and hence, improving performance by changing weights and the learning rate of the neural network. Finding the optimisation method suitable for the dataset can play a vital role when it comes to prediction accuracy. Some commonly used optimisers are gradient descent, stochastic gradient descent, momentum, nesterov accelerated gradient and adaptive moment estimation (ADAM). Every optimiser has its own strengths and weaknesses, for example, gradient descent is computationally inexpensive but requires large memory. On the other hand, ADAM converges rapidly but is computationally expensive \cite{optimisers}. 

Epoch is when a complete training data has once passed through the neural network and updated the internal model parameters. Most of the time all samples of the training data cannot pass in one go and hence, each epoch is divided into several batches. Single epoch may result in underfitting and multiple epochs are required as the internal parameters need to be changed multiple times for better generalisation. Once the number of epochs starts increasing the model starts fitting better until it reaches optimal fitting and if the number of epochs are still increased then the overfitting can happen \cite{afaq2020S}.

Batch size defines the number of samples that ANN algorithm must walk through before updating internal parameters. Batch size can affect the performance of ANN algorithm in terms of prediction accuracy and training time. In general, smaller batch size is better in generalising to the test set than large batch training \cite{samuel2017}. 

Activation function lets the neural network learn complex patterns in the data by adding non linearity into the model. Various non linear activation functions are available such as Sigmoid, Softmax, Tanh, ReLU etc. Choice of activation function can impact the performance of the training process \cite{hayou2019}. The activation function for input layer, hidden layer and output layer may be different based on the dataset.

Based on the domain knowledge, resource restrictions and dataset characteristics, a number of hyperparameter options are chosen. ANN optimiser tests the prediction accuracy of the neural network using these optimisers with a small subset of data to find the most suitable set of options.

\subsection{Explainability of our Model} \label{explainability}
Advanced predictive models like ANN achieve good performance by using complex models which are difficult to explain. However, assessing how a particular prediction is made is vital for understanding and trusting the model by the end user. We assess the explainability of our model using SHapley Additive exPlanations (SHAP) by identifying feature relevance on the basis of each sample \cite{Lundberg2017}. SHAP explains what the model is doing by finding the important relationships between the input features and the predicted outcome, which allows the users to interpret the predictions of our ANN model. SHAP uses an explainer method to combine the inputs to evaluate the effects on the predictive model and produces a locally interpretable model for every prediction \cite{Lundberg2017}. The explainer illustrates the SHAP value for each feature that illustrates the importance of each feature for particular prediction.  

After training the ANN model with a training subset of dataset, SHAP explainer analyses the testing subset. The explainer generates the output of the ANN model and assigns a SHAP value to each feature for prediction. These SHAP values indicate which feature increases or decreases the influence on the prediction to belong to a particular class \cite{Huang2020}. The summary plot is used to visualise the effects of the indicated features on prediction of the attack subcategory. We have used the force plot to show how features effect the prediction for a single value.

\section{Hyperparameter Selection by ANN Optimiser} \label{hyperparameterANN}
In this section we discuss the hyperparameter selection by ANN optimiser for IoTID20 dataset for attack detection at binary, category and subcategory level. 

\subsection{ANN Setup} \label{ANNsetup}
The optimal ANN setup is found by using a grid search, which completes an all-encompassing search over the hyperparameter’s space. The grid search parameters included: 
\begin{itemize}
\item Number of epochs 
\item Batch size
\item Number of neurons
\item optimiser function
\item Activation function for input and hidden layers
\item Activation function for output layer
\end{itemize}

It is worth mentioning that the experiments were conducted with one, two, or three hidden layers. The results were not significantly different. Therefore, in this study, only one hidden layer was chosen for hyperparameter selections.

The full cycle of learning and adapting the weights of the network is called epoch. In this model, 100 epochs and 200 epochs were used in the grid search. The particular count of samples utilised in one iteration is called batch size. 10 and 100 batches were used in this grid search. The units or nodes in the input and hidden layers are called neurons. In this model, 100 neurons and 200 neurons were used in the grid search. 

The optimiser functions evaluated in this model are: 
\begin{itemize}
\item Adaptive Moment Estimation (Adam) - a stochastic optimisation method that uses a randomly selected data subset to create a stochastic approximation, instead of using the entire data set to calculate the actual gradient. 
\item Root Mean Square Propagation (rmsprop) - an extension of gradient descent and the AdaGrad version of gradient descent that uses a decaying average of partial gradients in the adaptation of the step size for each parameter.
\item Stochastic gradient descent (SGD) -  an iterative method for optimising an objective function with suitable smoothness properties. It can be regarded as a stochastic approximation of gradient descent optimisation, since it replaces the actual gradient by an estimate thereof.
\item AdaMax  (AdaMax) - an optimisation method that implements the AdaMax algorithm. It is a variant of Adam based on the infinity norm.
\item AdaGrad (Adagrad) - a stochastic optimisation method that adapts the learning rate to the parameters. It performs smaller updates for parameters associated with frequently occurring features, and larger updates for parameters associated with infrequently occurring features.
\end{itemize}

In the input and hidden layer, and in the output layer, the activation functions were evaluated independently. The Activation functions evaluated in this model are: 
\begin{itemize}
\item Rectified Linear Unit (ReLU) - the range of the activation value is from 0 to infinity.
\item Hyperbolic Tangent (tanh) - the range of the activation value is from -1 to 1.
\item Sigmoid (sigmoid) - the range of the activation value is from 0 to 1.
\item Softplus (softplus) - it is the softer (smoother) version of ReLU and softplus. The range of the activation value is from 0 to infinity.
\item Softmax (softmax) - it is used for multiclass classification by converting a vector of values to a probability distribution. The elements of the output vector are in range (0, 1) and sum to 1.
\end{itemize}


The evaluations were conducted for three types of classifications: 
\begin{itemize}
\item Binary classification of normal and attack classes
\item Category level attack classification
\item Subcategory level attack classification
\end{itemize}


For each classification model, it required 1,000 experiments to evaluate all the pre-defined hyperparameter options. That is Epochs (2) $\times$ Batch (2) $\times$ Neurons (2) $\times$ optimiser (5) $\times$ Activation I (5) $\times$ Activation II (5) = 2 $\times$ 2 $\times$ 2 $\times$ 5 $\times$ 5 $\times$ 5 = 1000 experiments. The aim of the study is to show that the results of subset dataset can be effectively used to classify the full dataset. Therefore, this study required 6,000 experiments to include subset and full dataset experiments. The first 3,000 experiments were conducted over a subset of the full dataset to find the optimal hyperparameter selection. and the second 3,000 experiments were conducted to validate the optimal selection with the full dataset of 625,783 records.
\subsection{Dataset}
In this work we used IOTID20 dataset \cite{Ullah2020} as this dataset was collected in smart home environment which consists of multiple smart home devices.
The IOTID20 dataset contains total 87 features, 84 network features and three labels. For the feature selection, we first dropped unimportant features including `Flow\_ID', `Src\_IP', `Dst\_IP', `Dst\_Port', and `Protocol'. Next, features with only single values are removed, including `Fwd\_PSH\_Flags', `Fwd\_URG\_Flags', `Fwd\_Byts/b\_Avg', `Fwd\_Pkts/b\_Avg', `Fwd\_Blk\_Rate\_Avg', `Bwd\_Byts/b\_Avg', `Bwd\_Pkts/b\_Avg', `Bwd\_Blk\_Rate\_Avg', `Init\_Fwd\_Win\_Byts', and `Fwd\_Seg\_Size\_Min'.
For the binary label, there are 40,073 Normal records and 585,710 Anomaly records. For category labels, there are Normal (40,073), DoS (59,391), Mirai (415,677, MITM ARP Spoofing (35,377) and Scan (75,265). For subcategory labels, there are Normal (40,073), DoS(59391), MiraiAck Flooding (55124), Mirai Brute Force (121181), Mirai HTTP Flooding (55818), Mirai UDP Flooding (183,554), MITM (35377), Scan Host Port (22,192) and Scan Port OS (53073).


\subsection{Performance Metrics}
In this work, the accuracy, precision, recall, F1-score, and support are used as performance measures. It is worth noting that the macro and weighted versions of the measures are considered for imbalanced multi-class classification. The aim is to have higher accuracy, precision, recall, F1-score. The support is the number of true instances for each class.

\begin{equation}\label{eq:accuracy}
Accuracy=\sum_{i=1}^{n\_class}\frac{TP+TN}{TP+TN+FP+FN}/n\_class,
\end{equation}

\begin{equation}\label{eq:precision}
Precision_{macro}=\sum_{i=1}^{n\_class}\frac{TP_i}{TP_i+FP_i}/n\_class,
\end{equation}
\begin{equation}\label{eq:precision2}
Precision_{weighted}=\sum_{i=1}^{n\_class}(\frac{TP_i}{TP_i+FP_i}\frac{support_i}{\sum_{j=1}^{n\_class}support_j}),
\end{equation}

\begin{equation}\label{eq:recall}
Recall_{macro}=\sum_{i=1}^{n\_class}\frac{TP_i}{TP_i+FN_i}/n\_class,
\end{equation}
\begin{equation}\label{eq:recall2}
Recall_{weighted}=\sum_{i=1}^{n\_class}(\frac{TP_i}{TP_i+FN_i}\frac{support_i}{\sum_{j=1}^{n\_class}support_j}),
\end{equation}

\begin{equation}\label{eq:f1score}
\text{F1-score}_{macro}=\frac{2\times Precision_{macro}\times Recall_{macro}}{Precision_{macro}+Recall_{macro}},
\end{equation}
\begin{equation}\label{eq:f1score2}
\text{F1-score}_{weighted}=\frac{2\times Precision_{weighted}\times Recall_{weighted}}{Precision_{weighted}+Recall_{weighted}},
\end{equation}
where $n\_class$ is the number of classes, $TP_i$, $TN$, $FP$ and $FN$ are the number of true positives of the $i$th class, the number of true negatives of the $i$th class, the number of false positives of the $i$th class and the number of false negatives of the $i$th class, respectively.

\subsection{Optimal Hyperparameter Selection using Subset}
\begin{figure}
\begin{subfigure}{.33\textwidth}
  \centering
  \includegraphics[width=1.1\linewidth]{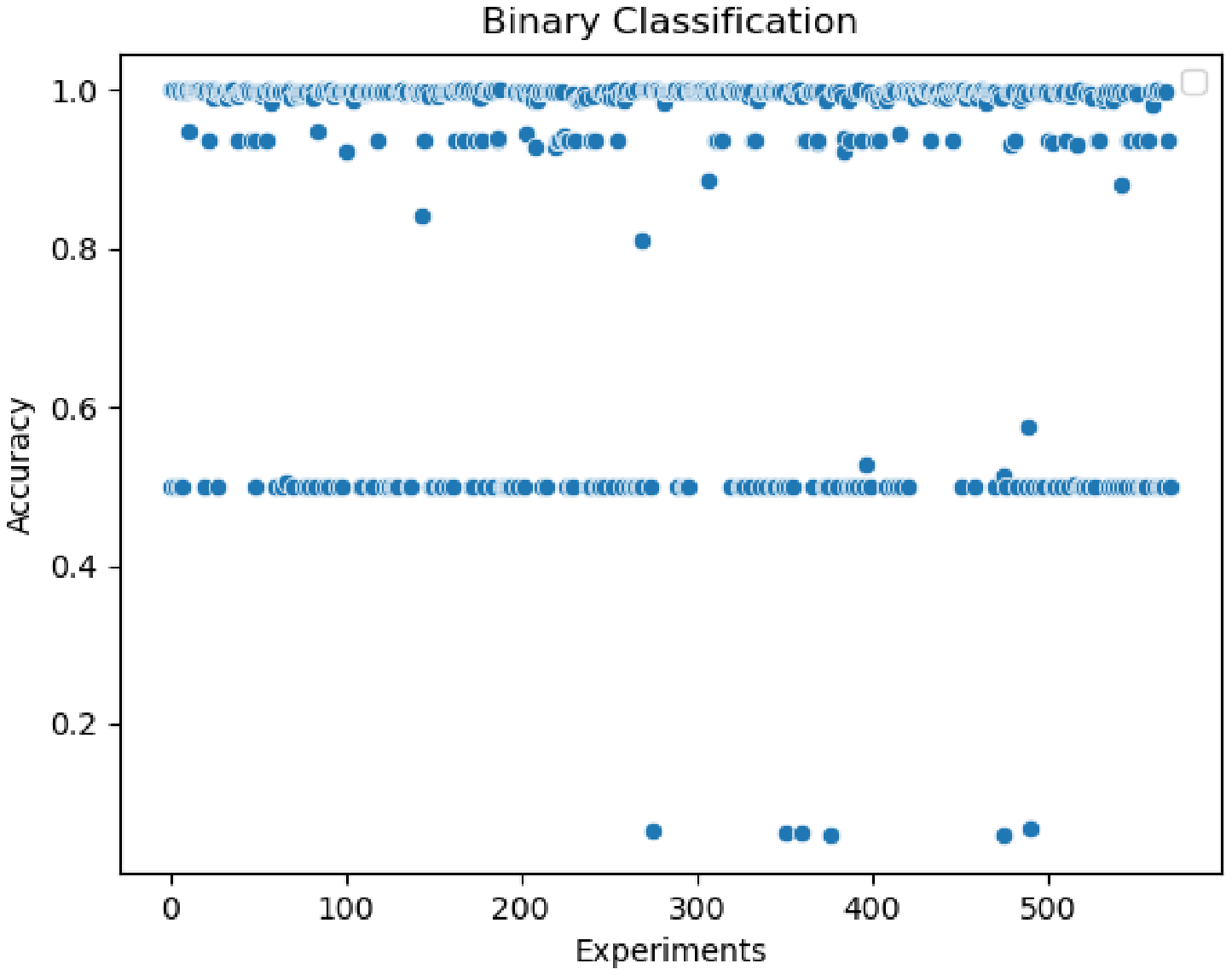}
  \caption{}
\end{subfigure}%
\begin{subfigure}{.33\textwidth}
  \centering
  \includegraphics[width=1.1\linewidth]{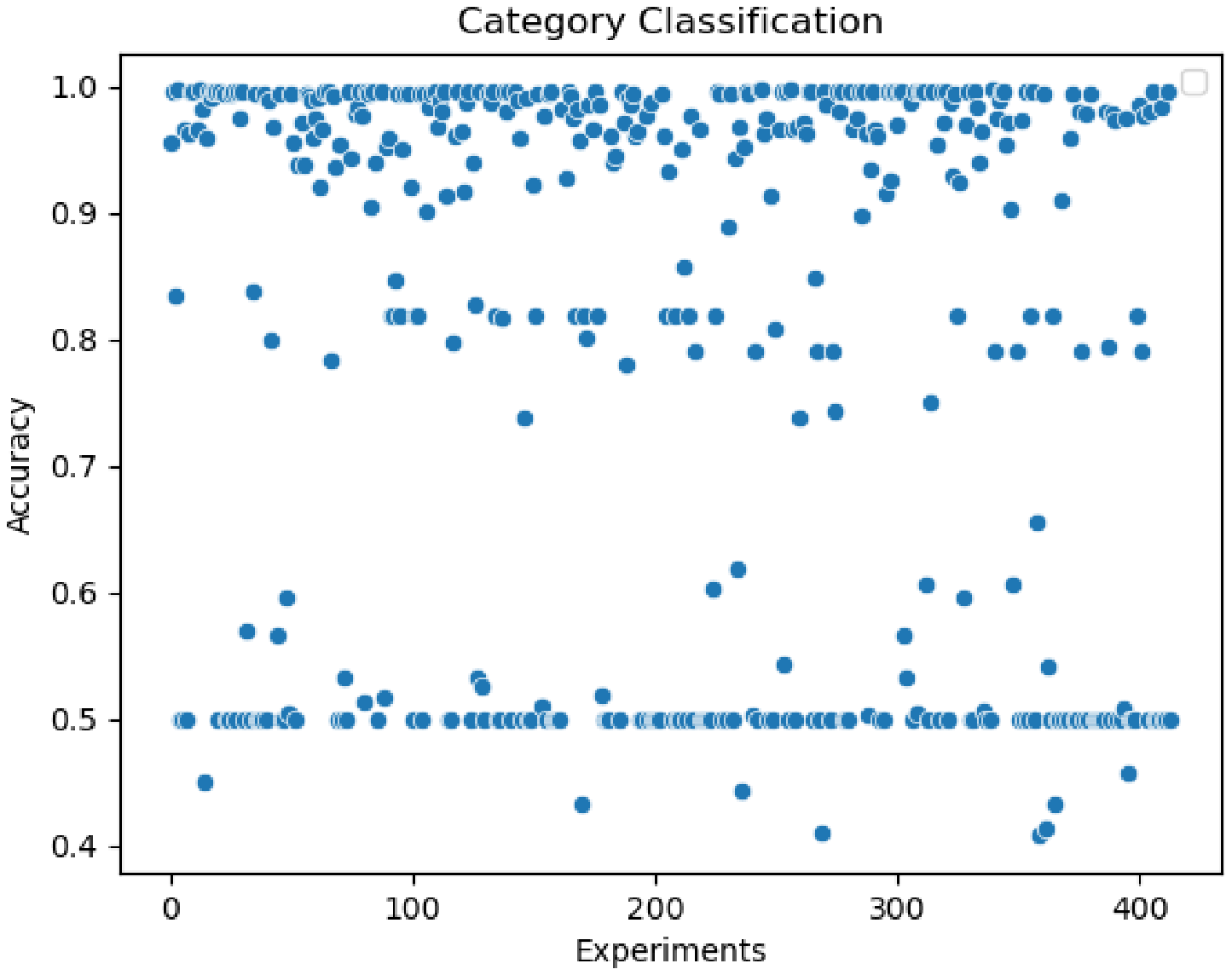}
  \caption{}
\end{subfigure}
\begin{subfigure}{.33\textwidth}
  \centering
  \includegraphics[width=1.1\linewidth]{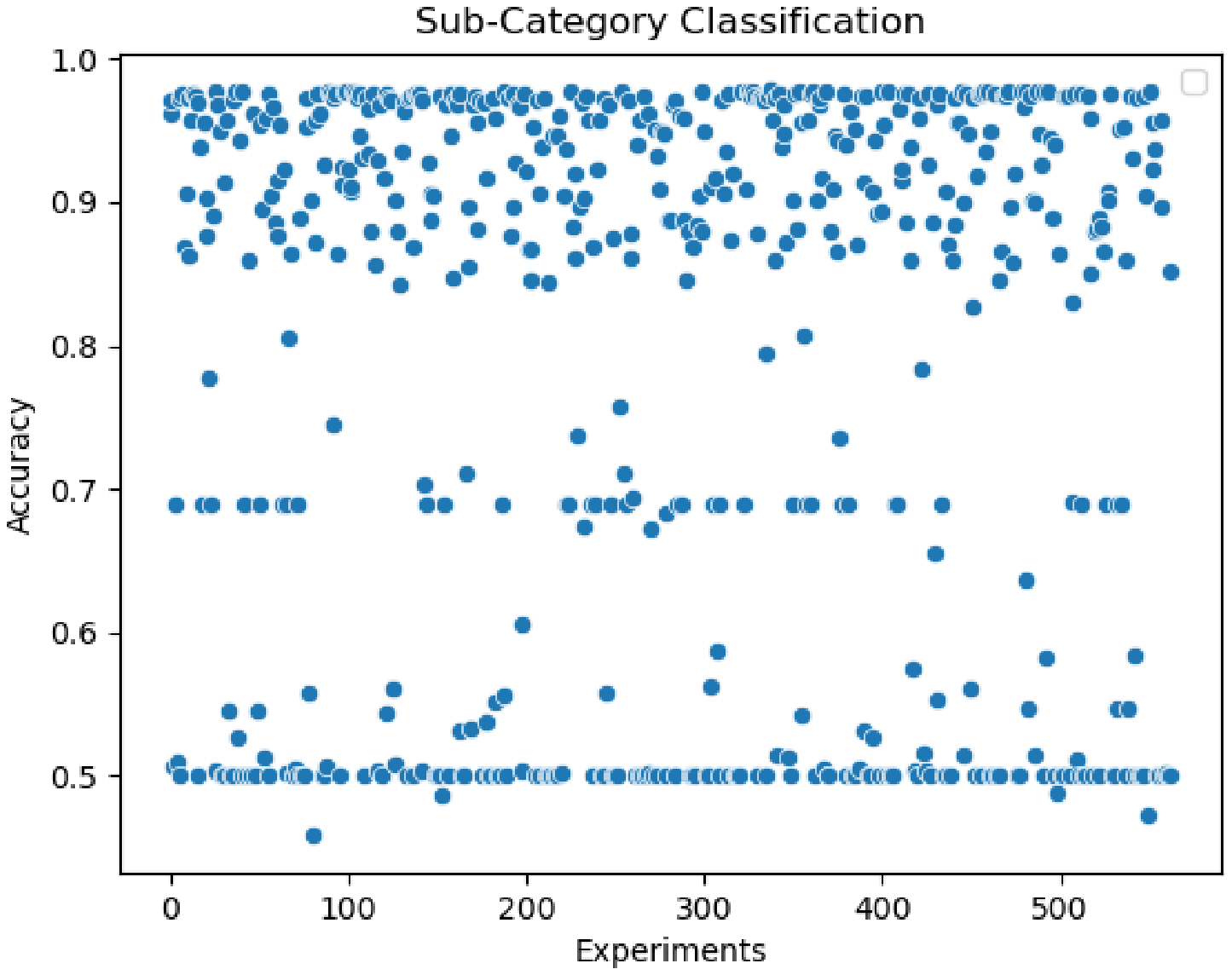}
  \caption{}
\end{subfigure}
\caption{Accuracy Results from (a) binary (b) category and (c) subcategory classification from Subset dataset experiments.}
\label{fig:subsetExperiments}
\end{figure}

Because of the mismatch data structure between the data and optimiser activation functions, some experiments were not successful. The analysis is based on the successful experiments over the subset with 10,000 records. 464 out of 1000 binary experiments, 474 out of 1000 category experiments, and 437 out of 1000 subcategory experiments returned successful results. The results over three different classifications were distributed on scatter plots in Figure \ref{fig:subsetExperiments}. By observing this figure, it is easy to see that, with the increasing number of target types, the accuracy rates become lower.

To understand the distribution of the high accuracy rates over 90\%, the summary of the results using the predefined hyperparameters for binary classification, category classification, and subcategory classification are listed in Table \ref{tab:hyperparameter}. As we can see from the table, it is much more challenging to reach 99\% with the number of prediction targets increases in the experiments. 57\% of binary experiments, 0.5\% of category experiments, and 0\% subcategory experiments reached over 99\% accuracy rates. 

\begin{table*}[htbp]
  \centering
\caption{Top 10 Best hyperparameter setups for binary classification using the full dataset.}
\label{tab:hyperparameter}
  \begin{tabular}{m{180pt}<{\centering} m{40pt}<{\centering} m{27pt}<{\centering} m{40pt}<{\centering}}
 \hline
Classification & Accuracy & Results & Percentage \\ \hline
\multirow{3}{*}{Binary (successful experiments = 464)}
& 99+\% & 263 & 57\% \\
& 95+\% & 372 & 80\% \\
& 90+\% & 434 & 93\% \\ \hline
\multirow{3}{*}{Category (successful experiments = 447)}
& 99+\% & 2 & 0.50\% \\
& 95+\% & 262 & 59\% \\
& 90+\% & 335 & 75\% \\ \hline
\multirow{3}{*}{Subcategory (successful experiments = 437)}
& 99+\% & 0 & 0\% \\
& 95+\% & 45 & 10\% \\
& 90+\% & 227 & 52\% \\ \hline
\end{tabular}
\end{table*}

To select the optimal hyperparameters, two approaches were taken in this study. The first one is to investigate the hyperparameter settings from the experiments with highest accuracy results (Approach I). The second approach is to investigate each hyperparameter setting together with its accuracy levels from all experiments (Approach II). 

Therefore, Approach I is to investigate the top 10 results of three classifications (Table \ref{tab:bestHyperparameter}). 

\begin{table*}[htbp]
\begin{center}
\caption{Top 10 Best hyperparameter setups for binary classification using full dataset.}
\label{tab:bestHyperparameter}
  \begin{tabular}{m{45pt}<{\centering} m{30pt}<{\centering} m{30pt}<{\centering} m{18pt}<{\centering} m{18pt}<{\centering} m{30pt}<{\centering} m{35pt}<{\centering} m{35pt}<{\centering}}
 \hline
Output & Accuracy & Neurons & Batch & Epoch & Optimiser & Activation I &
Activation II 
\\ \hline
Binary & 99.88\% & 100 & 10 & 200 & Adam & tanh & sigmoid \\
& 99.86\% & 100 & 100 & 200 & rmsprop & tanh & sigmoid \\
& 99.85\% & 200 & 100 & 200 & Adam & tanh & sigmoid \\
& 99.84\% & 200 & 10 & 100 & Adam & tanh & sigmoid \\
& 99.84\% & 100 & 10 & 100 & Adam & tanh & sigmoid \\
& 99.84\% & 200 & 100 & 100 & Adam & tanh & sigmoid \\
& 99.83\% & 100 & 100 & 200 & Adam & tanh & sigmoid \\
& 99.83\% & 200 & 100 & 200 & rmsprop & tanh & softmax \\
& 99.82\% & 200 & 100 & 100 & Adam & tanh & softmax \\
& 99.82\% & 200 & 100 & 200 & rmsprop & tanh & sigmoid \\ \hline
Category & 99.04\% & 200 & 10 & 200 & Adamax & tanh & softmax \\
& 99.03\% & 100 & 100 & 200 & Adam & tanh & softmax \\
& 99.00\% & 200 & 10 & 100 & Adam & sigmoid & softmax \\
& 98.98\% & 100 & 10 & 200 & Adam & sigmoid & softmax \\
& 98.96\% & 100 & 10 & 200 & AdaMax & tanh & softmax \\
& 98.93\% & 200 & 100 & 200 & Adam & tanh & softmax \\
& 98.93\% & 200 & 100 & 100 & Adam & tanh & softmax \\
& 98.92\% & 200 & 10 & 100 & Adam & tanh & softmax \\
& 98.92\% & 200 & 10 & 100 & AdaMax & tanh & softmax \\
& 98.89\% & 200 & 10 & 200 & AdaMax & ReLU & softmax \\ \hline
Subcategory & 95.65\% & 100 & 10 & 200 & AdaMax & tanh & softmax \\
& 95.63\% & 200 & 10 & 100 & Adam & sigmoid & softmax \\
& 95.62\% & 100 & 10 & 200 & Adam & sigmoid & softmax \\
& 95.61\% & 100 & 100 & 200 & Adam & tanh & softmax \\
& 95.58\% & 100 & 10 & 100 & Adam & tanh & softmax \\
& 95.56\% & 100 & 100 & 200 & rmsprop & tanh & softmax \\
& 95.55\% & 200 & 100 & 100 & Adam & tanh & softmax \\
& 95.53\% & 100 & 10 & 100 & Adam & sigmoid & softmax \\
& 95.50\% & 200 & 10 & 200 & AdaMax & tanh & softmax \\
& 95.49\% & 100 & 10 & 100 & rmsprop & tanh & softmax \\ \hline
\end{tabular}
\end{center}
\end{table*}

Table \ref{tab:bestHyperparameter} shows the range of top 10 accuracy for binary is between 99.82-99.88\%, Category is between 98.89-99.04\%, and Subcategory is between 95.49-95.65\%. It is obvious to see the optimal options for Optimiser, Activation I, and Activation II. The optimal optimiser for top 10 classification results is \textbf{adam}, which ran 21 (7+6+6) out of 30 experiments. The optimal Activation I function for top 10 classification results is \textbf{tanh}, which conducted 24 (10+7+7) out of 30 experiments. The optimal Activation II function for top 10 classification results is \textbf{softmax}, which was used in 22 (2+10+10) out of 30 experiments. However, it is difficult to identify the optimal options for the number of Neuron, Batch, and Epochs. For each hyperparameter, a similar number of either option was used for top 10 classifications.

Approach II is to investigate how well the hyperparameters were used in all experiments in terms of accuracy. Six bar charts were generated to show the accuracy levels of each hyperparameter option for three classification types in all successful experiments.

\begin{figure}
\begin{subfigure}{.33\textwidth}
  \centering
  \includegraphics[width=1.3\linewidth]{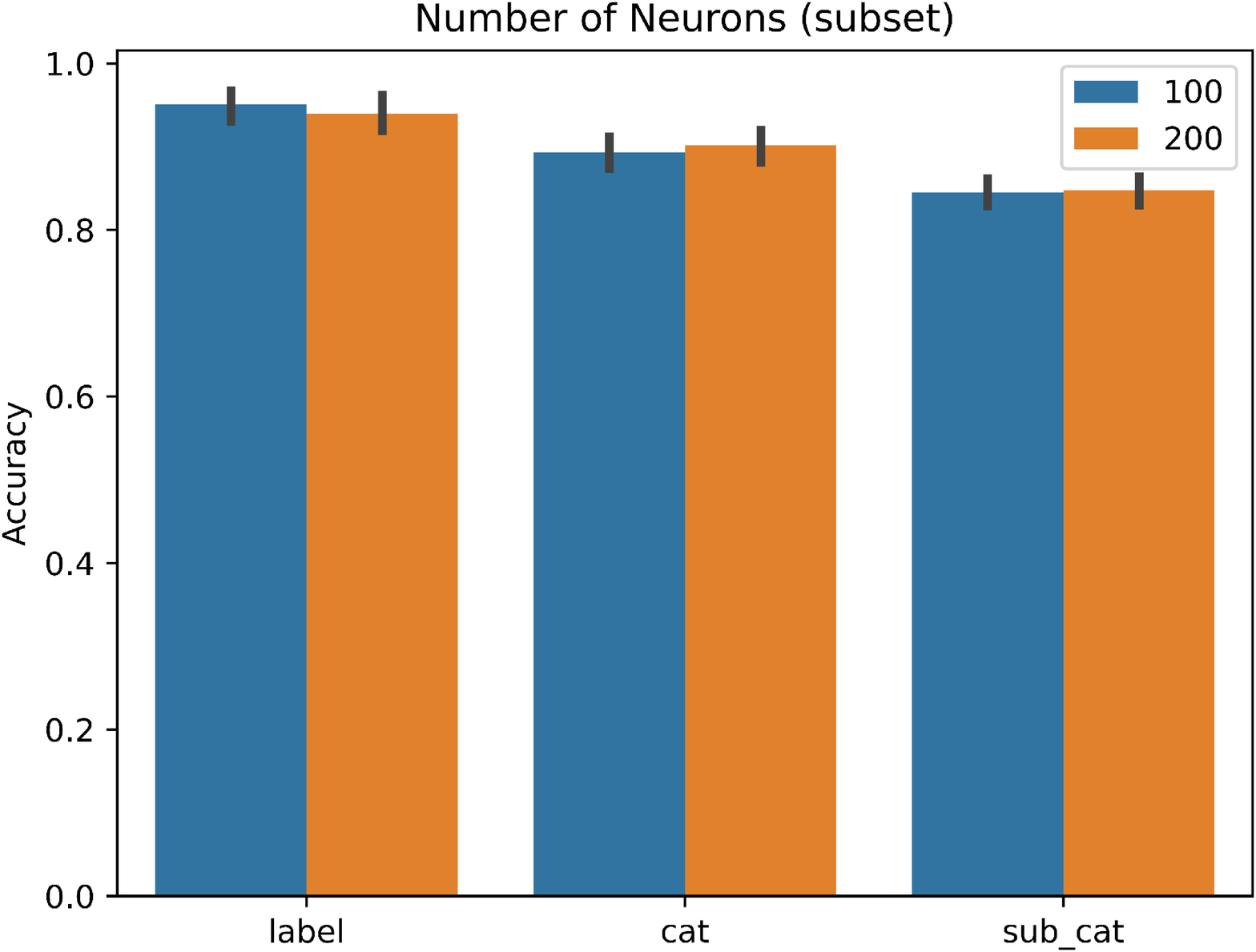}
  \caption{Number of neurons.}
\end{subfigure}%
\begin{subfigure}{.33\textwidth}
  \centering
  \includegraphics[width=1.3\linewidth]{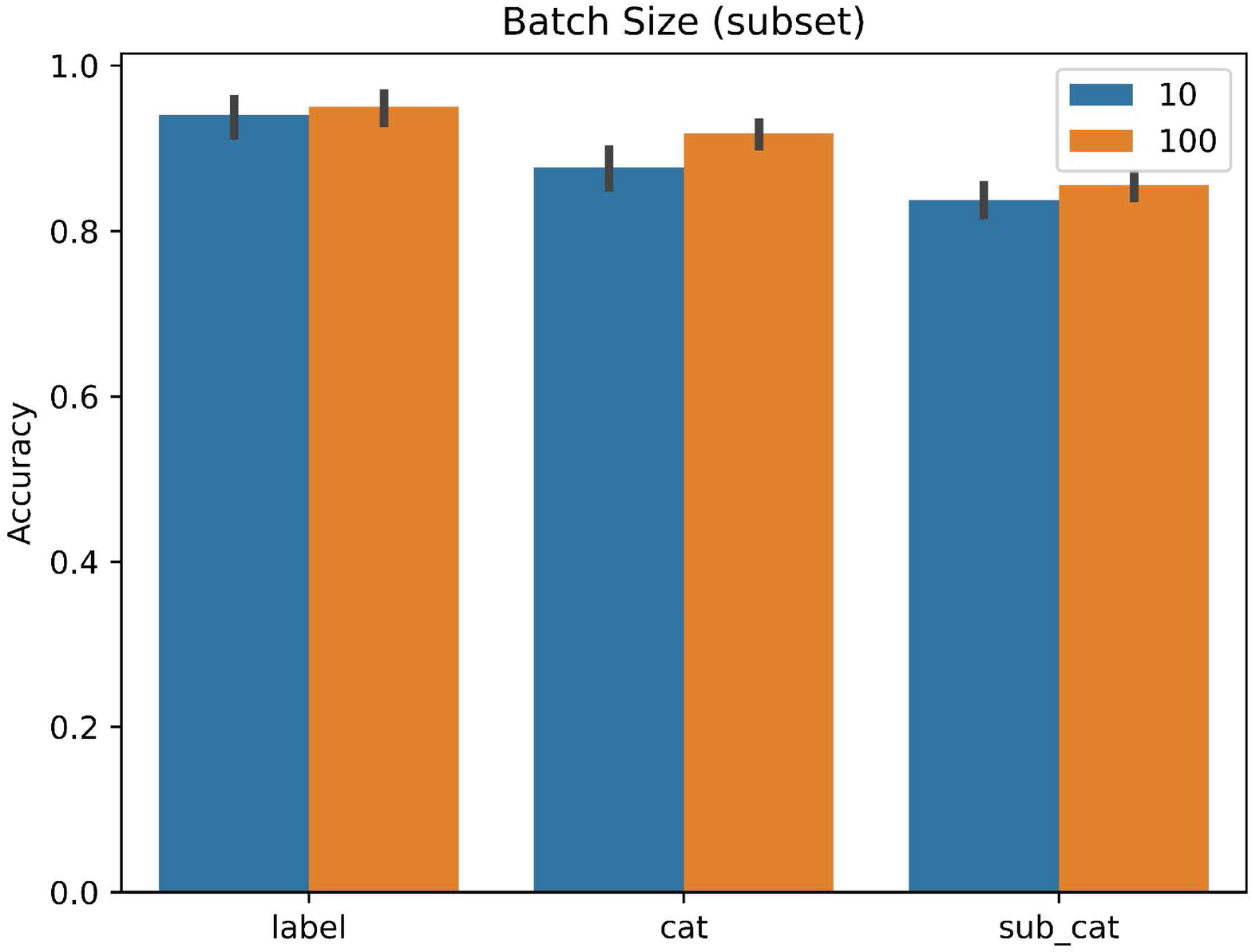}
  \caption{Batch size.}
\end{subfigure}
\begin{subfigure}{.33\textwidth}
  \centering
  \includegraphics[width=1.3\linewidth]{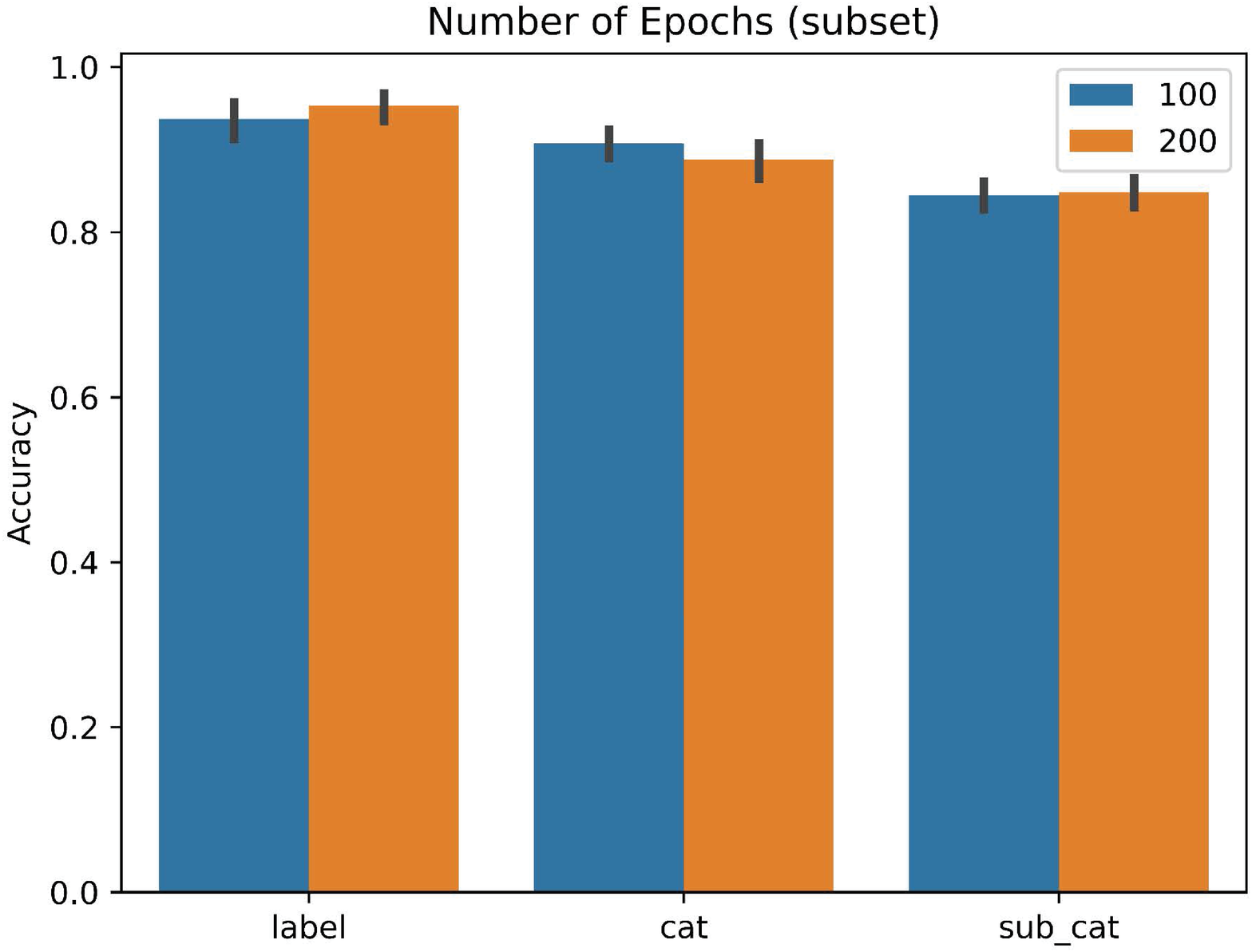}
  \caption{Number of epochs.}
\end{subfigure}
\begin{subfigure}{.33\textwidth}
  \centering
  \includegraphics[width=1.3\linewidth]{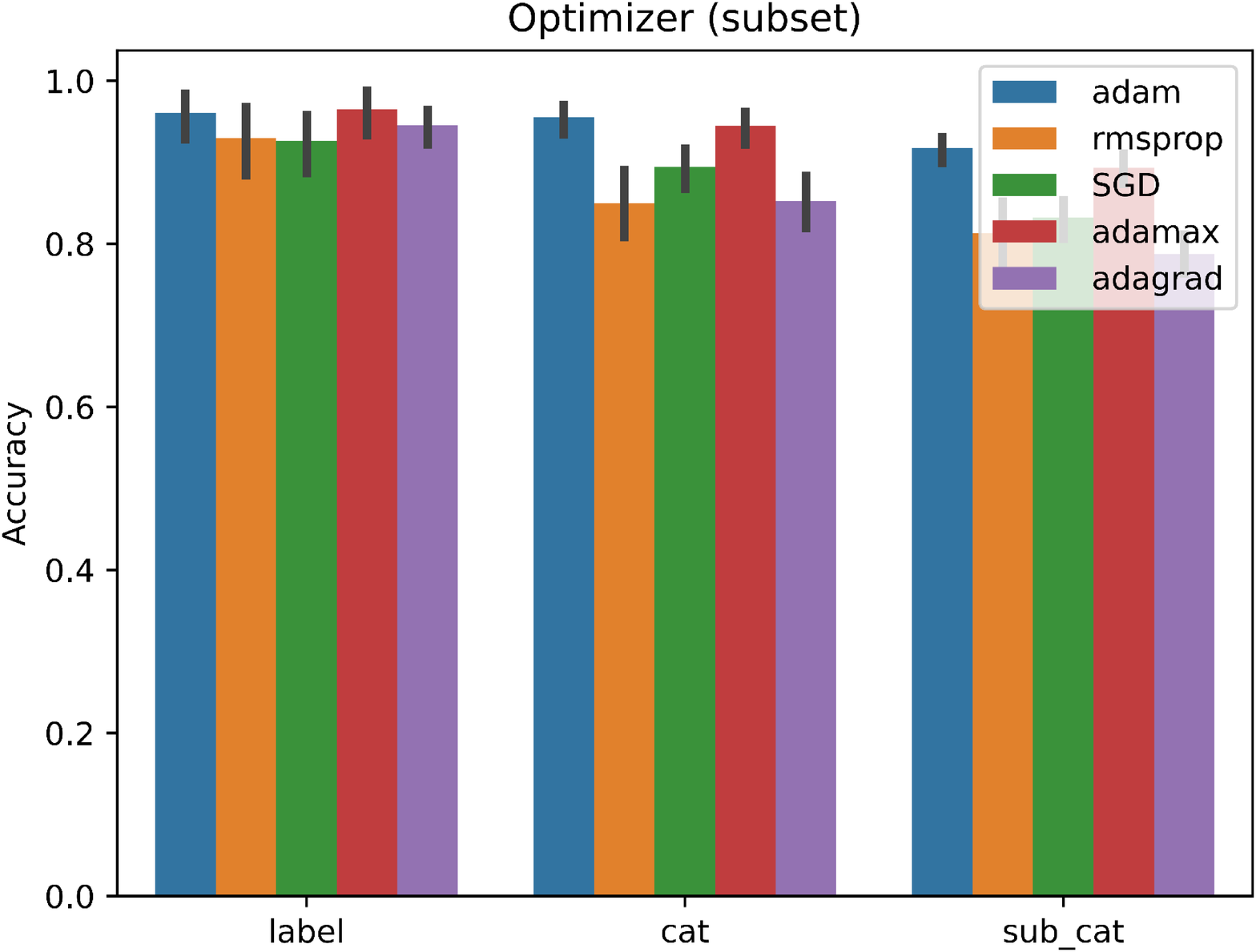}
  \caption{Optimiser.}
\end{subfigure}%
\begin{subfigure}{.33\textwidth}
  \centering
  \includegraphics[width=1.3\linewidth]{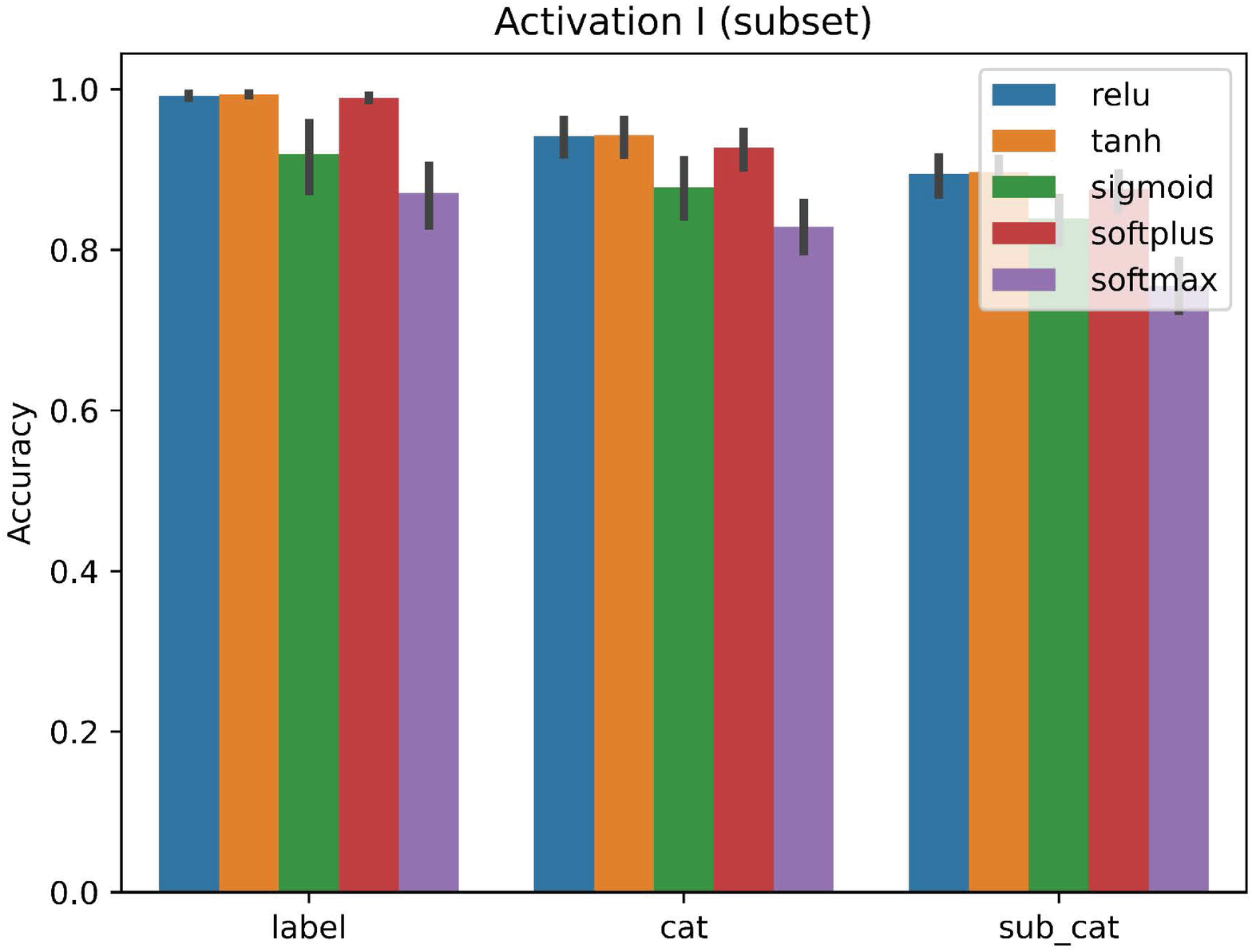}
  \caption{Activation I.}
\end{subfigure}
\begin{subfigure}{.33\textwidth}
  \centering
  \includegraphics[width=1.3\linewidth]{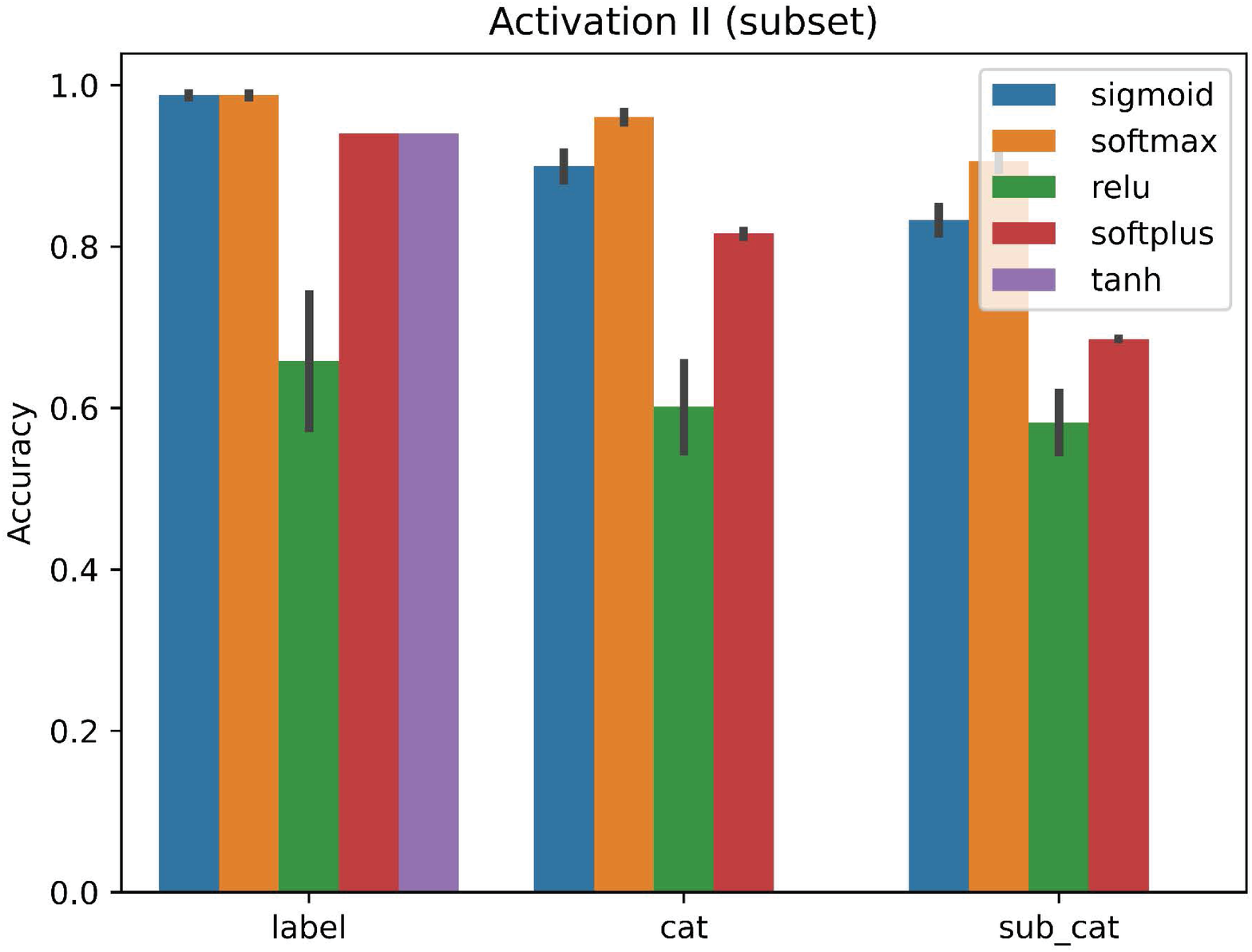}
  \caption{Activation II.}
\end{subfigure}
\caption{Different hypterparameters for ANN.}
\label{fig:hyperparameters}
\end{figure}

The bar charts of epochs, batch and neurons show not much difference in either hyperparameter selections as shown in Figure \ref{fig:hyperparameters}. The results from this approach are similar to Approach I. Combining the results from both approaches, the optimal selections were made. 200 was selected as the optimal number of neurons, 100 was selected as the batch size, and 200 was selected as the optimal number of epochs.

Using Approach I, \textbf{adam} was selected for optimiser, \textbf{tanh} was selected for Activation I, and \textbf{softmax} was selected for Activation II. This selection was supported by the results from Approach II. In Fig. \ref{fig:hyperparameters}, \textbf{adam}, \textbf{tanh}, and \textbf{softmax} generated the highest accuracy as optimiser, Activation I, and Activation II. The final selection of the hyperparameters is shown in (Table \ref{tab:optimalParameters}).

\begin{table*}[htbp]
\begin{center}
\caption{Selected hyperparameters based on Approaches I and II results}
\label{tab:optimalParameters}
\begin{tabular}{cc}
 \hline
Hyperparameter & Selection \\ \hline
Epochs & 200 \\
Batch & 100 \\
Neurons & 200 \\
Optimiser & adam \\
Activation I & tanh \\
Activation II & softmax \\ \hline
\end{tabular}
\end{center}
\end{table*}

\subsection{Validation with Full Dataset}
Although the full dataset was used for validation purposes, full experiments were conducted on this dataset to support the optimal hyperparameter selection from the subset dataset. The procedures are explained in the following steps. Through the grid search, 3,000 experiments were conducted on the full dataset with 625,783 records.

\subsubsection{Binary Classification of Normal and Attack Classes} 
\label{BC}
\begin{figure}
\begin{subfigure}{.33\textwidth}
  \centering
  \includegraphics[width=1.1\linewidth]{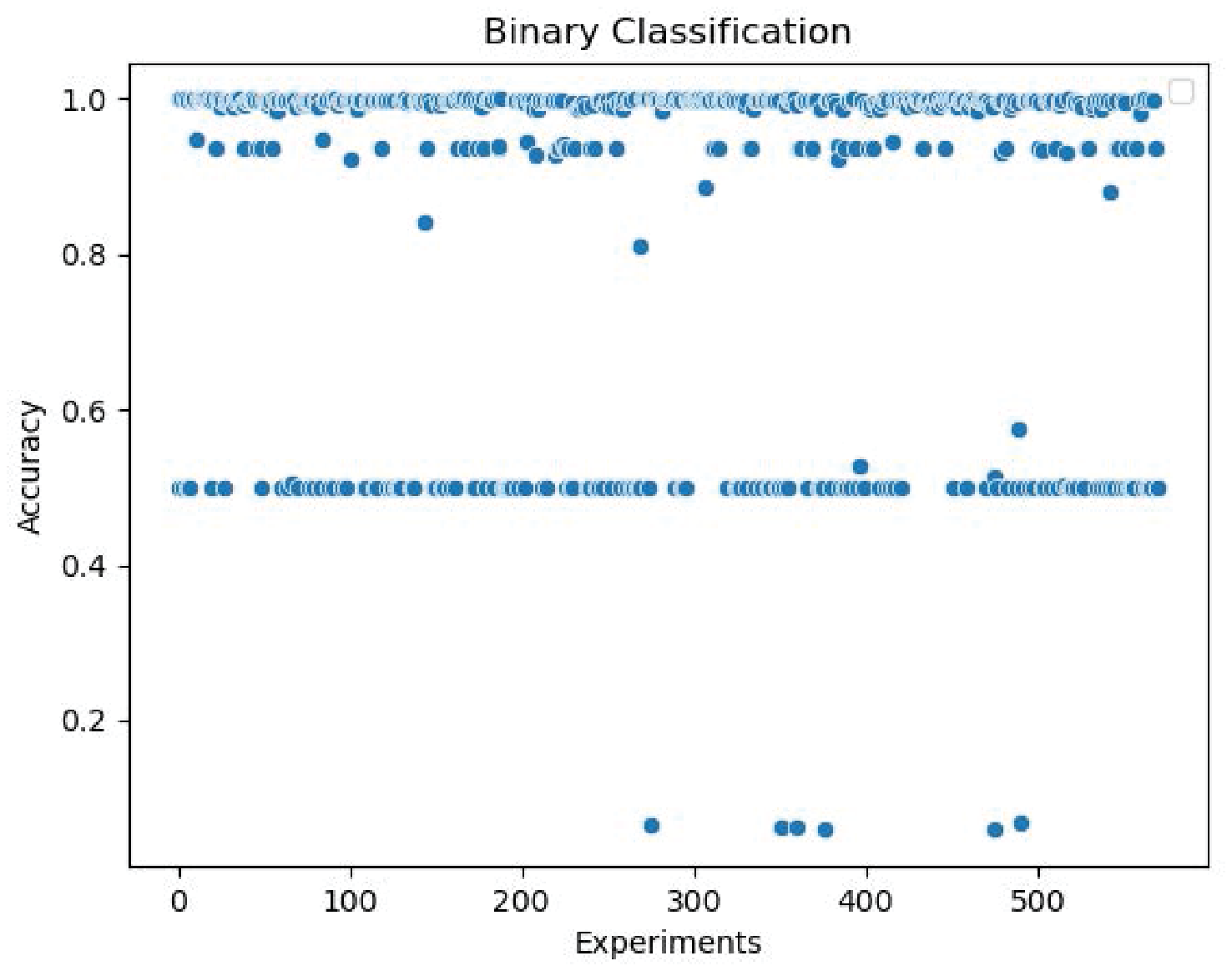}
  \caption{}
  \label{fig:4a}
\end{subfigure}%
\begin{subfigure}{.33\textwidth}
  \centering
  \includegraphics[width=1.1\linewidth]{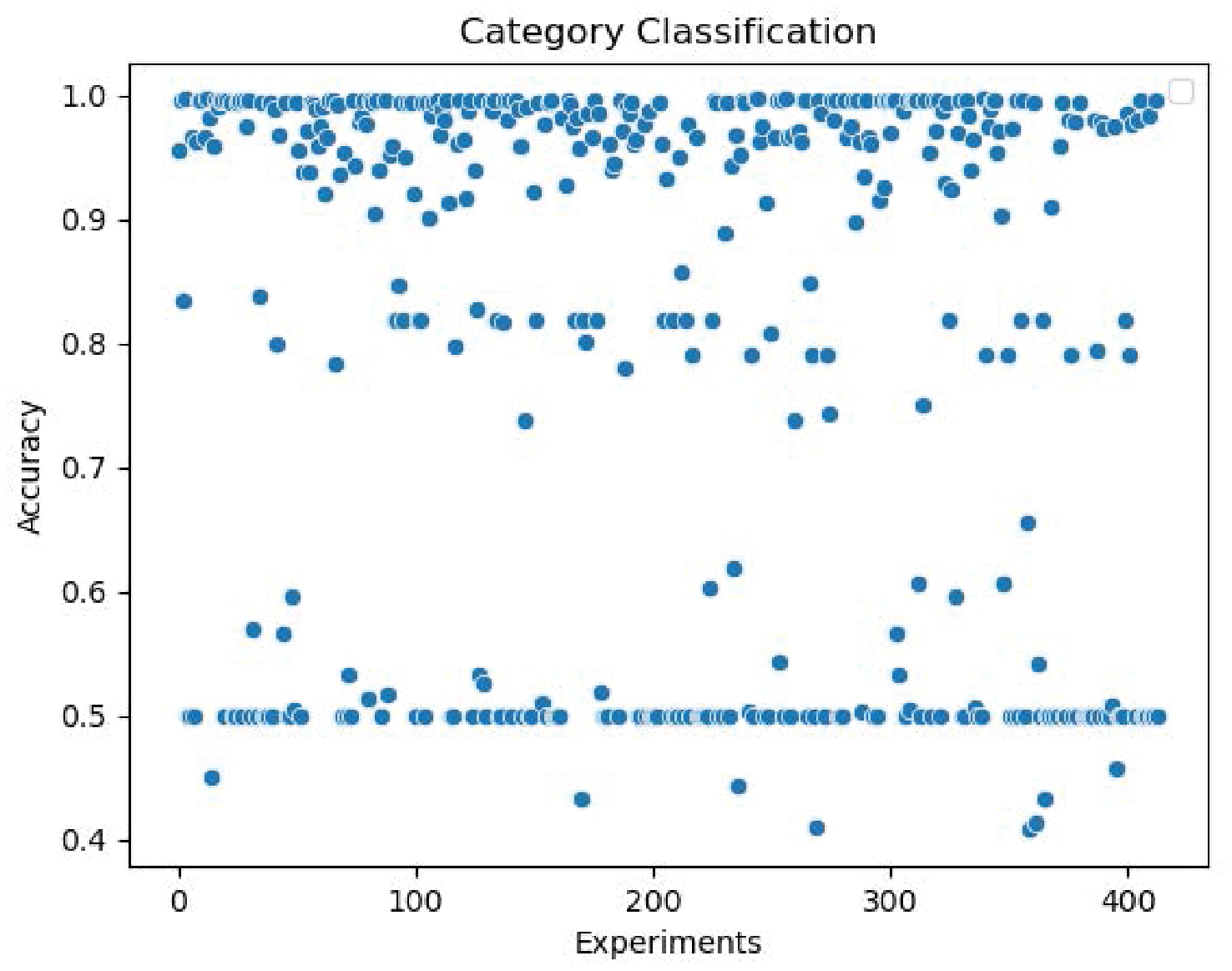}
  \caption{}
  \label{fig:4b}
\end{subfigure}
\begin{subfigure}{.33\textwidth}
  \centering
  \includegraphics[width=1.1\linewidth]{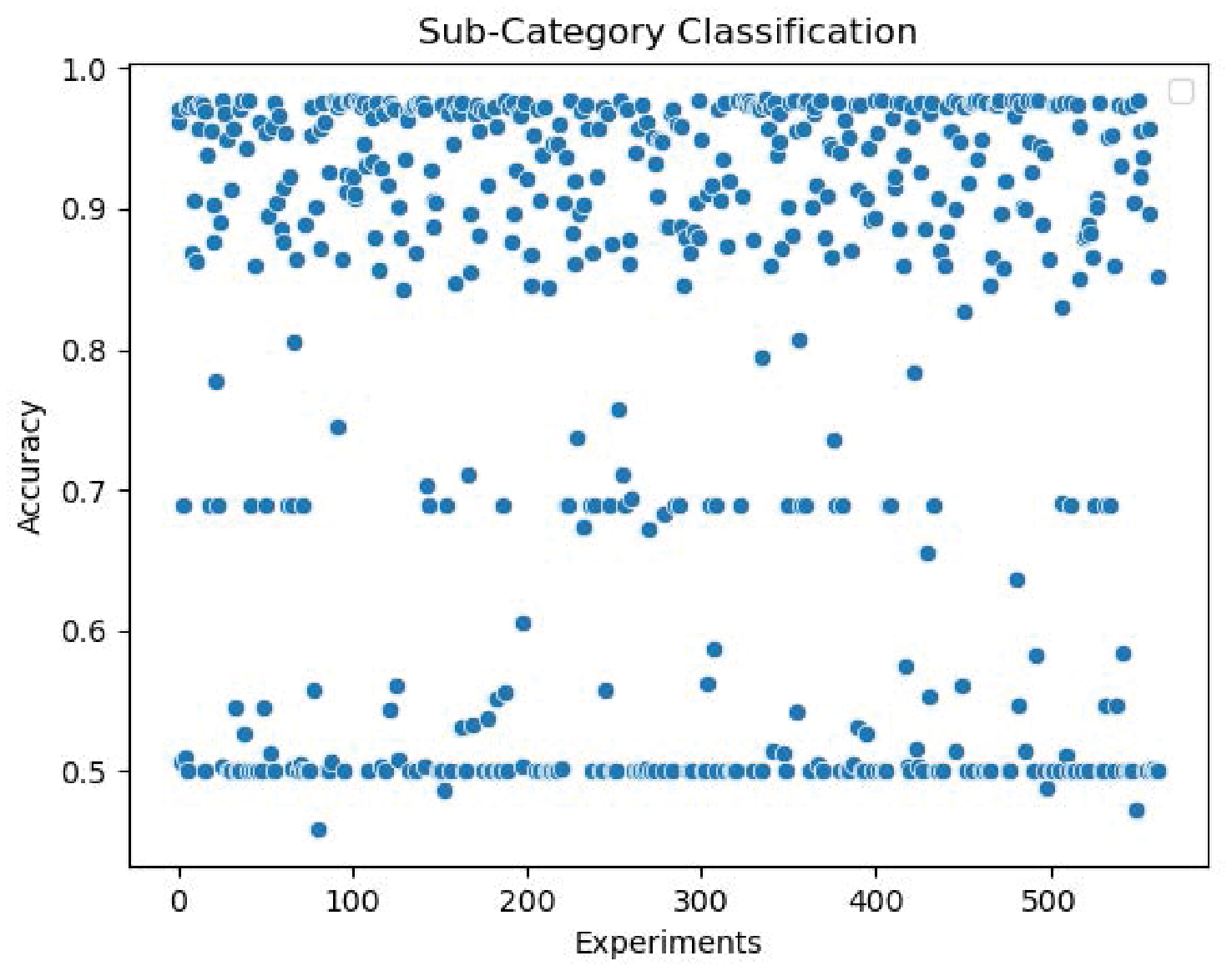}
  \caption{}
  \label{fig:4c}
\end{subfigure}
\caption{Accuracy Results from (a) binary (b) category and (c) subcategory classification from full dataset experiments.}
\label{fig:fulldataset}
\end{figure}

Among the 1000 experiments based on the grid search technique, 571 results successfully returned using the predefined hyperparameters. 74\% of experiments (421 out of 571) returned the results above 90\% accuracy. 63\% of experiments (361 out of 571) returned the accuracy above 95\%. 58\% of experiments 334 out of 571) returned above 99\%. The accuracy results from the 459 experiments were presented as a scatter plot shown in Figure \ref{fig:4a}.


The top 10 accuracy results with detailed hyperparameter setups on this binary classification are presented in Table \ref{tab:binaryParameter}. For all top 10 results, the accuracy rates stay between 99.951\% and 99.956\%. It is no surprising that all the Activation II (the output layer) is sigmoid because this is a classification problem. The best accuracy is 99.9559\%, with a mix of 200 neurons, AdaMax optimiser, ReLU activation function in hidden layers, batch size of 10, and 200 epochs. The activation function is ReLU in the top 5 accuracy results. The batch size of 10 is suitable for SGD optimiser while 100 for AdaMax.

\begin{table*}[htbp]
\begin{center}
\caption{Top 10 Best hyperparameter setups for binary classification using full dataset.}
  \begin{tabular}{m{45pt}<{\centering} m{30pt}<{\centering} m{18pt}<{\centering} m{18pt}<{\centering} m{30pt}<{\centering} m{55pt}<{\centering} m{58pt}<{\centering}}
 \hline
Accuracy & Neurons & Batch & Epoch & Optimiser & Activation I &
Activation II 
\\ \hline
0.999559 & 200 & 100 & 200 & AdaMax & ReLU & sigmoid\\
0.999553 & 200 & 10 & 200 & SGD & ReLU & sigmoid\\
0.999550 & 200 & 100 & 100 & AdaMax & ReLU & sigmoid\\
0.999540 & 100 & 10 & 200 & SGD & ReLU & sigmoid\\
0.999530 & 100 & 100 & 200 & AdaMax & ReLU & sigmoid\\
0.999521 & 200 & 100 & 100 & adam & sigmoid & sigmoid\\
0.999519 & 200 & 10 & 200 & SGD & tanh & sigmoid\\
0.999515 & 100 & 100 & 100 & adam & tanh & sigmoid\\
0.999515 & 100 & 100 & 200 & adam & sigmoid & sigmoid\\
0.999511 & 100 & 10 & 100 & SGD & ReLU & sigmoid\\ \hline
\end{tabular}
  \label{tab:binaryParameter}%
  \end{center}
\end{table*}

\subsubsection {Category Level Attack Classification} \label{CC}
Using the predefined hyperparameters, the grid search returned 414 results out of 1000 experiments. 53\% of experiments (221 out of 414) returned the results above 90\% accuracy. 47\% of experiments (193 out of 414) returned the accuracy above 95\%. 25\% of experiments (103 out of 414) returned the accuracy above 99\%. The accuracy results from the 449 experiments were presented as a scatter plot in Fig. \ref{fig:4b}. 


The top 10 accuracy results with detailed hyperparameter setups on this category classification (5 classes) are presented in Table \ref{tab:categoryParameter}. For all top 10 results, the accuracy rates stay between 99.7\% and 99.72\%, which are slightly worse than the binary classification (compared to Table \ref{tab:binaryParameter}). This is because with the increase of classes, it is more difficult to accurately predict the correct class. It is expected that for multi-class classification, the Activation II in the output layer is softmax activation function. The best accuracy is 99.7238\%, with a mix of 100 epochs, 200 neurons, adam optimiser, tanh activation function in hidden layers and batch size of 100. Among all the optimisers, adam has the best performance in the top 10 accuracy results.

\begin{table*}[htbp]
\begin{center}
\caption{Top 10 best hyperparameter setups for category classification full dataset.}
  \begin{tabular}{m{45pt}<{\centering} m{30pt}<{\centering} m{18pt}<{\centering} m{18pt}<{\centering} m{30pt}<{\centering} m{55pt}<{\centering} m{58pt}<{\centering}}
 \hline
Accuracy & Neurons & Batch & Epoch & Optimiser & Activation I &
Activation II 
\\ \hline
0.997238 & 200 & 100 & 100 & adam & tanh & softmax\\
0.997212 & 100 & 100 & 200 & adam & ReLU & softmax\\
0.997194 & 200 & 100 & 200 & adam & ReLU & softmax\\
0.997151 & 200 & 10 & 200 & AdaMax & tanh & softmax\\
0.997080 & 200 & 100 & 100 & adam & ReLU & softmax\\
0.997070 & 100 & 100 & 200 & adam & tanh & softmax\\
0.997041 & 100 & 10 & 200 & SGD & ReLU & softmax\\
0.997036 & 200 & 10 & 100 & AdaMax & tanh & softmax\\
0.997035 & 200 & 100 & 200 & adam & sigmoid & softmax\\
0.997027 & 100 & 100 & 200 & adam & sigmoid & softmax\\
\hline
\end{tabular}
  \label{tab:categoryParameter}%
  \end{center}
\end{table*}

\subsubsection{Subcategory Level Attack Classification} \label{SCC}
Using the predefined hyperparameters, the grid search returned 563 results out of 1000 experiments. 46\% of experiments (259 out of 563) returned the results above 90\% accuracy. 31\% of experiments (174 out of 563) returned the accuracy above 95\%. No experiments returned above 99\%. The accuracy results from the 563 experiments were presented as a scatter plot in Fig. \ref{fig:4c}. 


The top 10 accuracy results with detailed hyperparameter setups on this subcategory classification (9 classes) are presented in Table \ref{tab:subCatParameter}. For all top 10 results, the accuracy rates stay between 97.75\% and 97.84\%. The top 10 results are worse than the category and binary classification (compared to Tables \ref{tab:binaryParameter} and \ref{tab:categoryParameter}) because with the increase of classes, it is more difficult to accurately predict the correct class. For multi-class classification, the Activation II in the output layer is softmax activation function. The best accuracy is 99.84\%, with a mix of 200 epochs, 200 neurons, adam optimiser, sigmoid activation function in hidden layers and batch size of 100. The epoch for all the top 10 accuracy is 200. Among all the optimisers, adam has the best performance in the top 4 accuracy results.

\begin{table*}[htbp]
\begin{center}
\caption{Accuracy Results from subcategory classification from full dataset experiments.}
  \begin{tabular}{m{45pt}<{\centering} m{30pt}<{\centering} m{18pt}<{\centering} m{18pt}<{\centering} m{30pt}<{\centering} m{55pt}<{\centering} m{58pt}<{\centering}}
 \hline
Accuracy & Neurons & Batch & Epoch & optimiser & Activation I &
Activation II
\\ \hline
0.978400 & 200 & 100 & 200 & adam & sigmoid & softmax\\
0.978192 & 100 & 100 & 200 & adam & ReLU & softmax\\
0.978087 & 200 & 100 & 200 & adam & ReLU & softmax\\
0.977948 & 200 & 100 & 200 & adam & tanh & softmax\\
0.977947 & 200 & 10 & 200 & SGD & tanh & softmax\\
0.977938 & 200 & 100 & 200 & AdaMax & tanh & softmax\\
0.977841 & 200 & 100 & 200 & AdaMax & tanh & softmax\\
0.977635 & 100 & 10 & 200 & SGD & ReLU & softmax\\
0.977635 & 200 & 10 & 200 & AdaMax & tanh & softmax\\
0.977576 & 100 & 100 & 200 & adam & tanh & softmax\\
\hline
\end{tabular}
  \label{tab:subCatParameter}%
  \end{center}
\end{table*}

\subsection{Optimal hyperparameter selection}  
The accuracy results of different hyperparameter settings are compared and presented in Figure \ref{fig:hyperparameters_full}. The accuracy results are grouped by the classification types: Label (binary classification), Category(5-class classification), and Subcategory (9-class classification). 

\subsubsection{Number of Epochs}
\begin{figure}
\begin{subfigure}{.33\textwidth}
  \centering
  \includegraphics[width=1.3\linewidth]{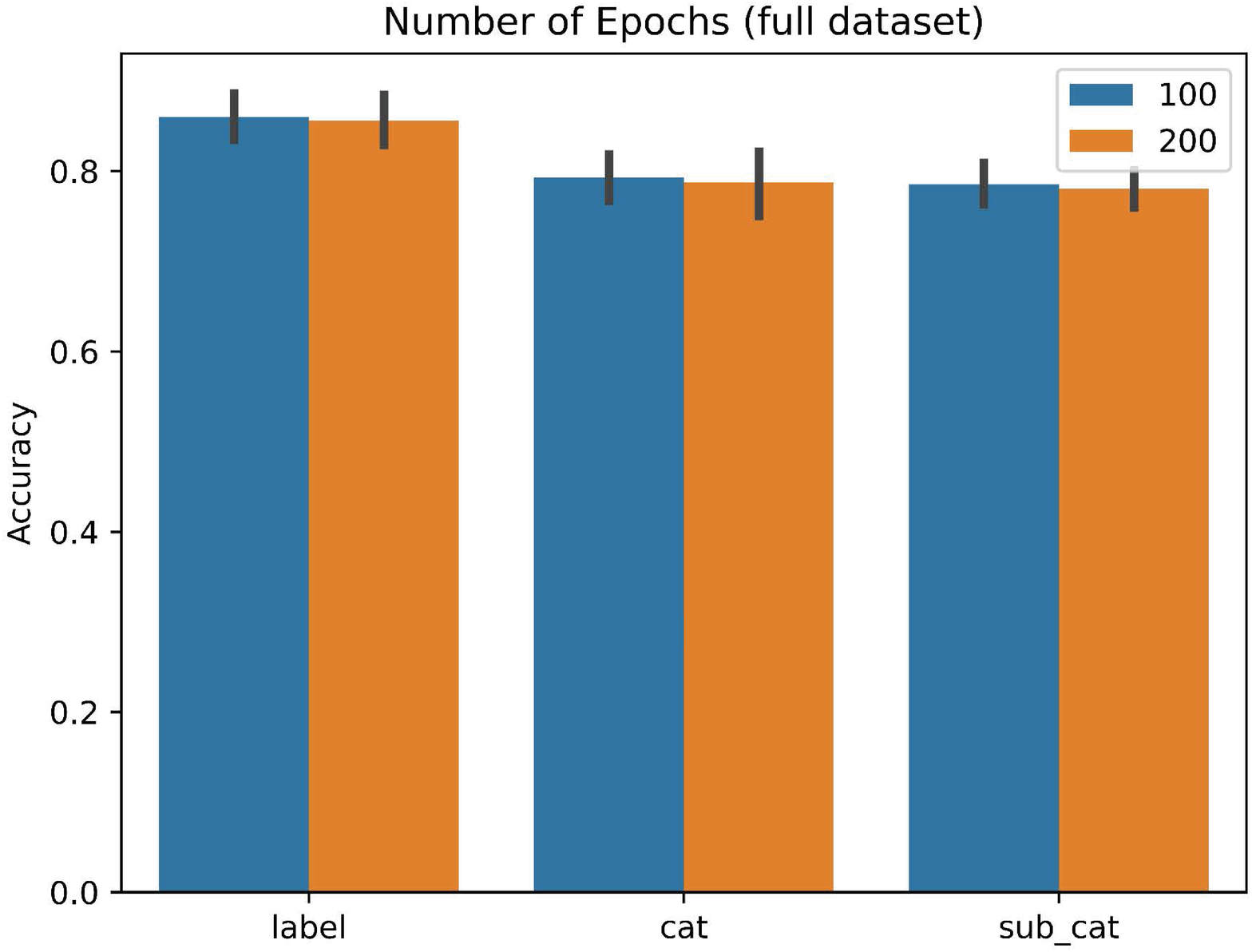}
  \caption{Number of Epochs}
\end{subfigure}%
\begin{subfigure}{.33\textwidth}
  \centering
  \includegraphics[width=1.3\linewidth]{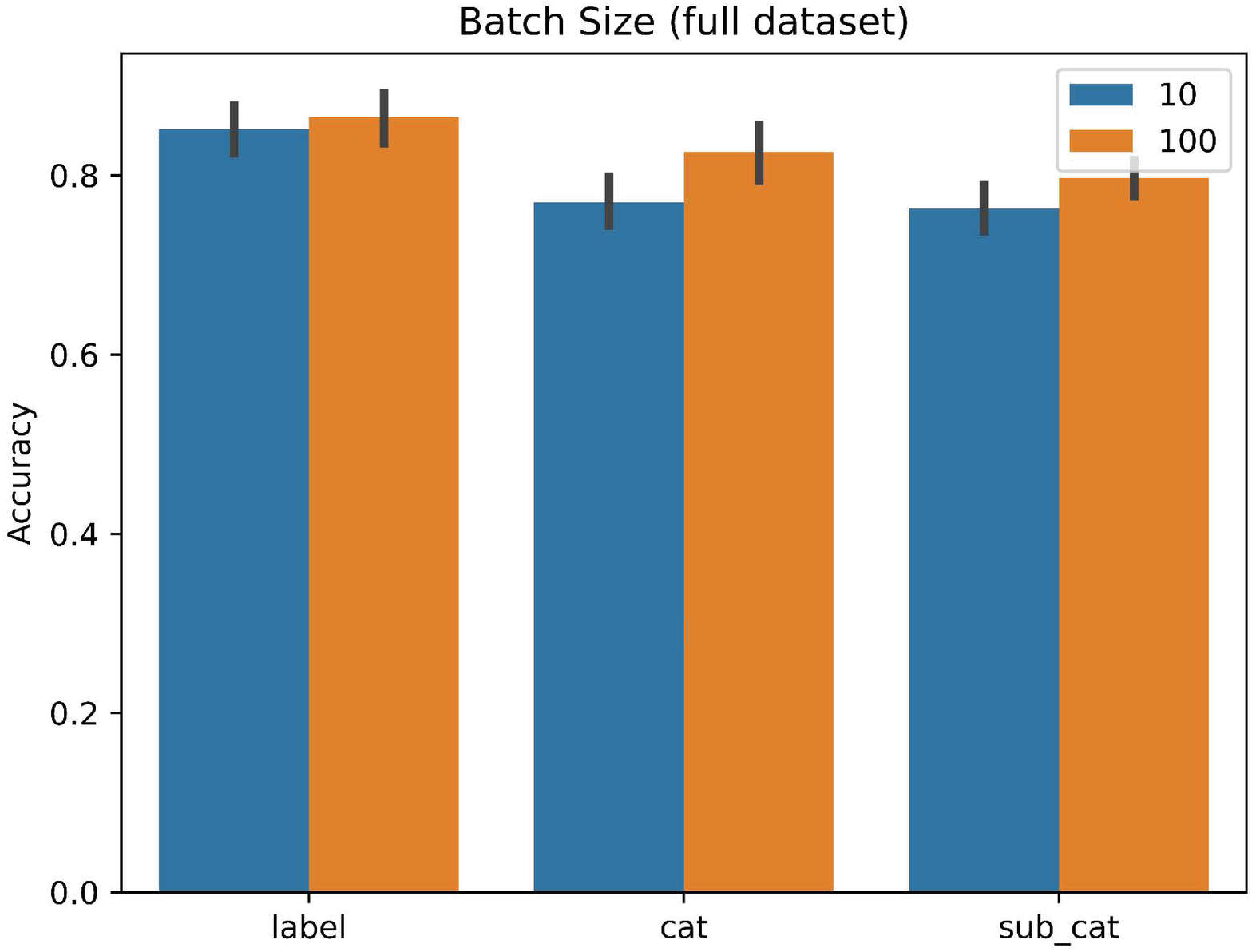}
  \caption{Batch size.}
\end{subfigure}
\begin{subfigure}{.33\textwidth}
  \centering
  \includegraphics[width=1.3\linewidth]{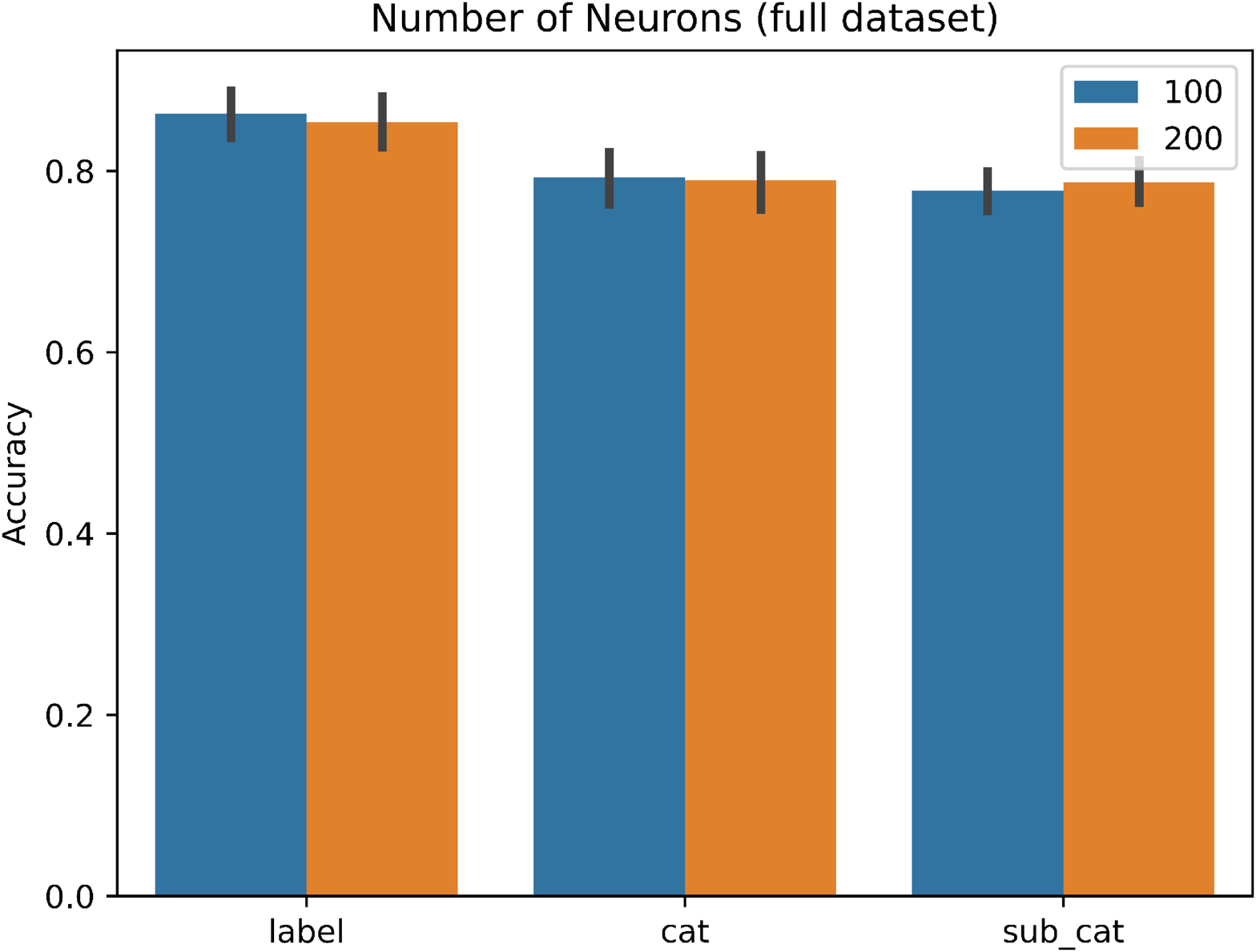}
  \caption{Number of Neurons.}
\end{subfigure}
\begin{subfigure}{.33\textwidth}
  \centering
  \includegraphics[width=1.3\linewidth]{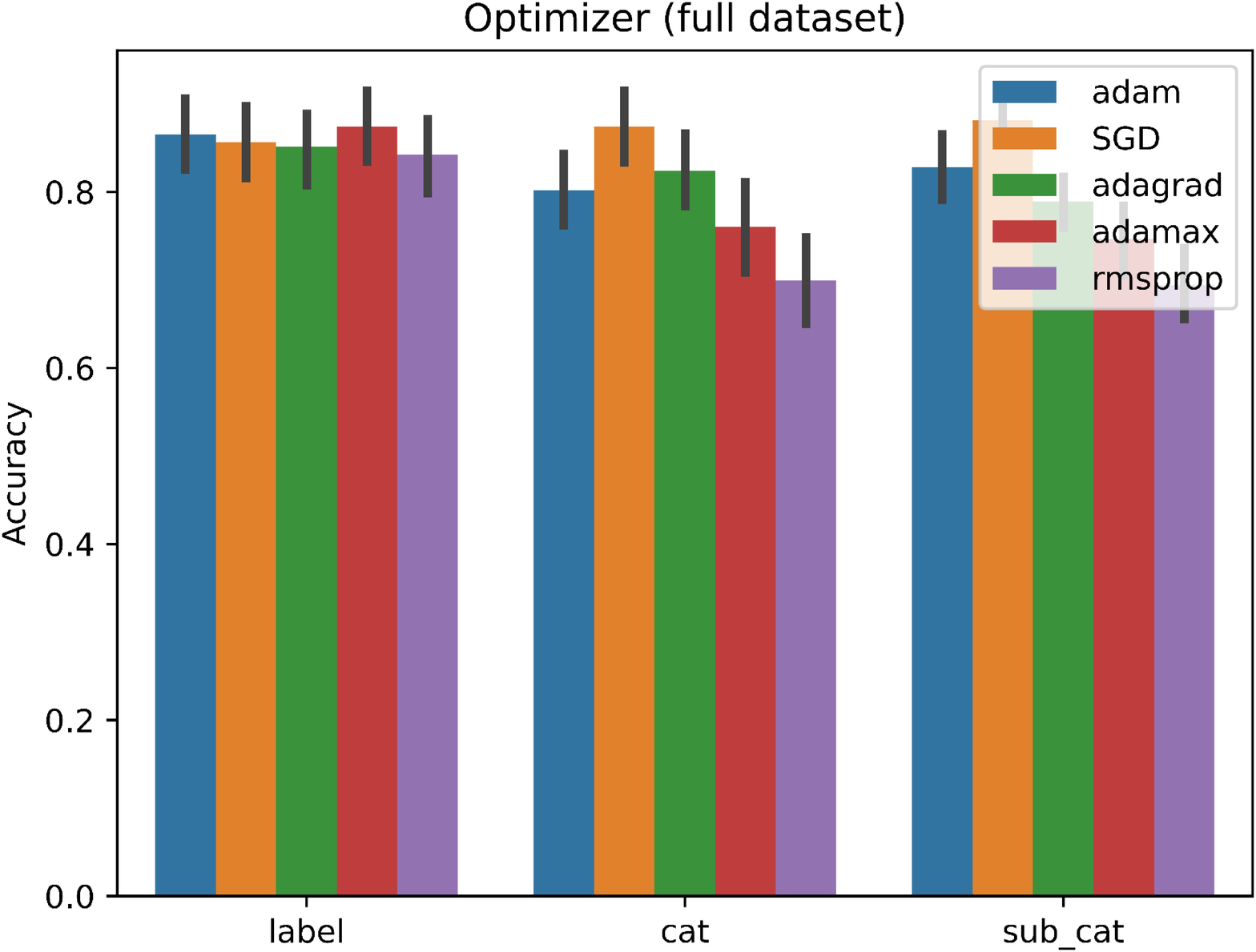}
  \caption{Optimiser.}
\end{subfigure}%
\begin{subfigure}{.33\textwidth}
  \centering
  \includegraphics[width=1.3\linewidth]{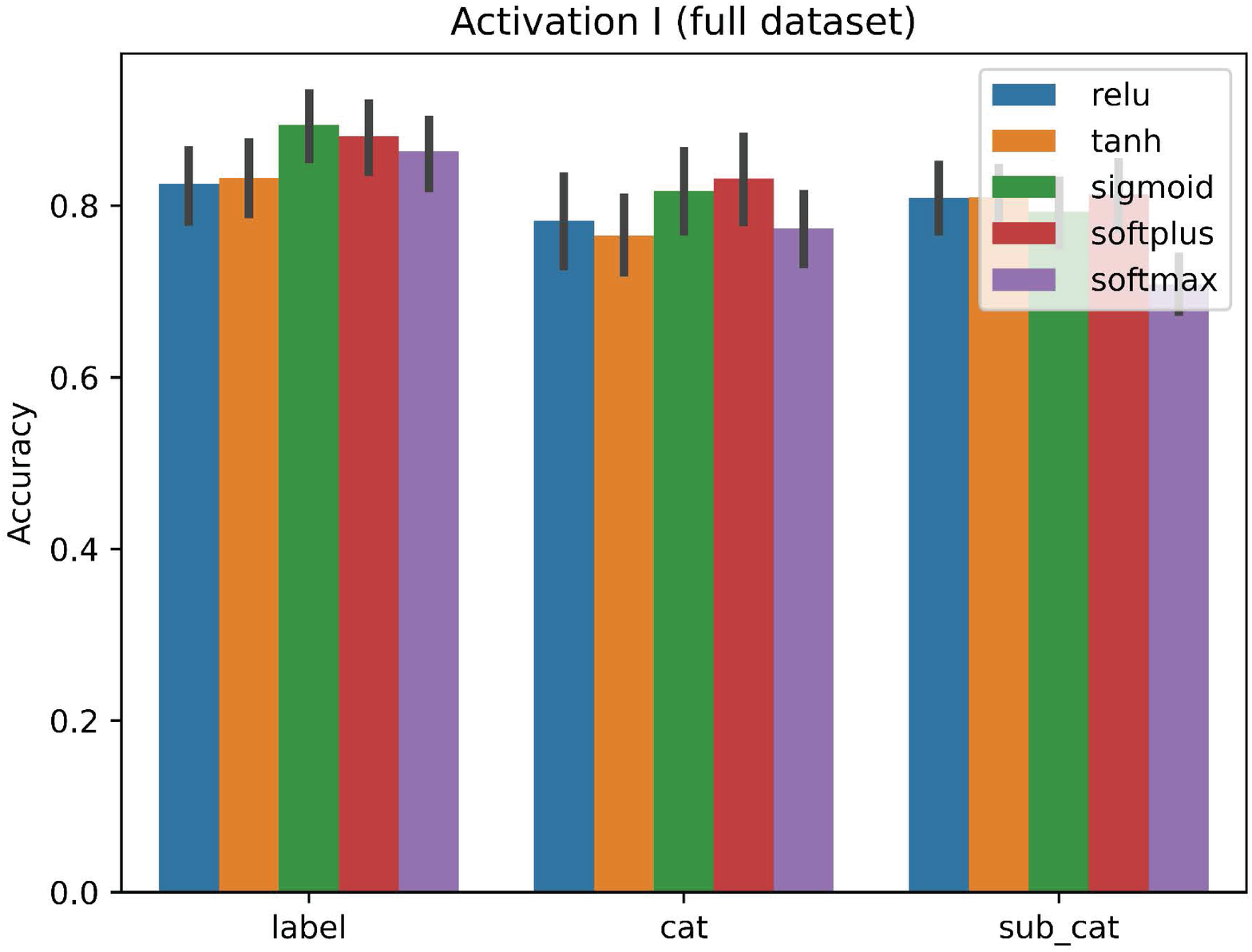}
  \caption{Activation I.}
\end{subfigure}
\begin{subfigure}{.33\textwidth}
  \centering
  \includegraphics[width=1.3\linewidth]{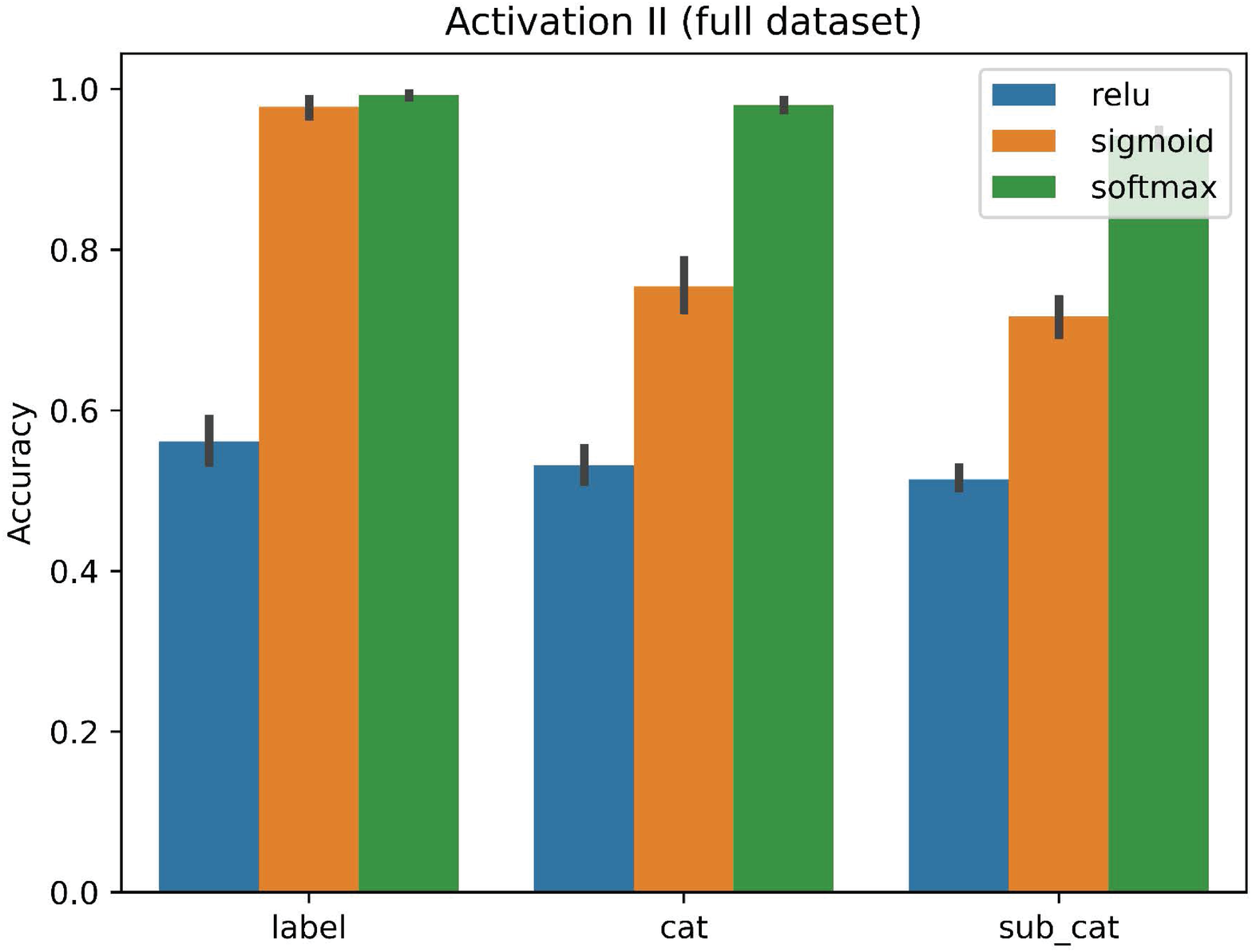}
  \caption{Activation II.}
\end{subfigure}
\caption{Accuracy for Three Classification Types with Different Hyperparameters}
\label{fig:hyperparameters_full}
\end{figure}

The accuracy rates of different predefined number of epochs are presented on a bar chart in Figure \ref{fig:hyperparameters_full}. Observing the results, there was almost no difference in choosing either 100 or 200 epochs. However, the top 10 accuracy results listed in Tables \ref{tab:binaryParameter} - \ref{tab:subCatParameter} show that 200 epochs provides better results for all three classification types. There were 6 out of 10 top 10 best hyperparameter settings used 200 epochs in label classification (Table \ref{tab:binaryParameter}), 7 out of 10 in category classification (Table \ref{tab:categoryParameter}), and 10 out of 10 in subcategory classification (Table \ref{tab:subCatParameter}).

\subsubsection{Batch Size}
The accuracy rates obtained by different batch size are presented on a bar chart in Figure \ref{fig:hyperparameters_full}. The results from this figure show that 100 as batch size provided high accuracy in all prediction experiments. The top 10 accuracy results listed in  Tables \ref{tab:binaryParameter} -  \ref{tab:subCatParameter} also present that 100 epochs provides better results for all three classification types. There were 6 out of 10 top 10 best hyperparameter settings used 200 epochs in label classification (Table \ref{tab:binaryParameter}), 7 out of 10 in category classification (Table \ref{tab:categoryParameter}), and 7 out of 10 in subcategory classification (Table \ref{tab:subCatParameter}). 

\subsubsection{Number of Neurons}
The accuracy rates by the number of neurons are presented on a bar chart Figure \ref{fig:hyperparameters_full}. From this figure, we can see that there was almost no difference in choosing either 100 or 200 neurons. However, the top 10 accuracy results listed in  Tables \ref{tab:binaryParameter} - \ref{tab:subCatParameter} indicate that 200 epochs may provide better results for category and subcategory types. 5 out of 10 top 10 best hyperparameter settings used 200 epochs in label classification (Table \ref{tab:binaryParameter}), 6 out of 10 in category classification (Table \ref{tab:categoryParameter}), and 8 out of 10 in subcategory classification (Table \ref{tab:subCatParameter}). 

\subsubsection{Optimiser}
The accuracy rates by five optimiser functions are presented on a bar chart Figure \ref{fig:hyperparameters_full}. By observing this figure, SGD could be the best hyperparameter option for all three classification types. In the top 10 accuracy results, AdaMax, adam, and SGD were used in the experiments. In label classification (Table \ref{tab:binaryParameter}), 3 out of top 10 accuracy results used AdaMax, 3 used adam, and 4 used SGD. In category classification (Table \ref{tab:categoryParameter}), 7 used adam, 3 used AdaMax, and 1 used SGD. In subcategory classification (Table \ref{tab:subCatParameter}), 5 used adam, 3 used AdaMax, and 2 used SGD. 

\subsubsection{Activation for Input and Hidden Layers}
The accuracy rates by five activation functions for input and hidden layers are presented on a bar chart Figure \ref{fig:hyperparameters_full}. No particular activation function can be observed as the best hyperparameter option for all three classification types. Relu, Tahn, and Sigmoid were selected by the top 10 accuracy results as the activation function for the input and hidden layers. In label classification (Table \ref{tab:binaryParameter}), 6 out of 10 top accuracy results used ReLU, 2 used tanh, and 2 used sigmoid. In category classification (Table \ref{tab:categoryParameter}), 4 used ReLU, 4 used tanh, and 2 used sigmoid. In sub\_category classification (Table \ref{tab:subCatParameter}), 3 used ReLU, 6 used tanh, and 1 used sigmoid. tanh may not be the best option for all classifications. It definitely shows the advantages in subcategory classification.

\subsubsection{Activation for the Output Layer}
The accuracy rates by five activation functions for output layers are presented on a bar chart Figure \ref{fig:hyperparameters_full}. As we can see from this figure, Tanh and Softplus are not the suitable algorithm for the data struction from this full dataset. Only ReLU, Sigmoid, and Softmax completed the computation and provided results. Based on this figure, Softmax could be the best hyperparameter option for the accuracy function for the output layer. By observing the top 10 accuracy results tables, softmax was used in 10 out of 10 experiments in Table \ref{tab:categoryParameter} and Table \ref{tab:subCatParameter}. Although Sigmoid was used as the activation function for the output layer in the top 10 accuracy results for Label classification in Table \ref{tab:binaryParameter}, softmax is the optimal activation function for category and subcategory classification. 

\section{Results} 
\label{results}
In this section we provide complete experimental results for subset dataset and full dataset for all three type of classification using the optimal set of hyperparameters.

\subsection{Binary Classification of Normal and Attack Classes}
Tables \ref{tab:binarySubDataset} and \ref{tab:binaryFullDataset} show the binary classification based on the subset dataset and full dataset, respectively.  In Table \ref{tab:binarySubDataset}, the f1-scores for macro f1-score, weighted f1-score, anomaly f1-score, and normal f1-score are 92.3214\%, 98.313\%, 99.1304\%, and 85.5124\%, respectively for subset of dataset.   In Table \ref{tab:categoryFullDataset}, the f1-scores for macro f1-score, weighted f1-score, anomaly f1-score, and normal f1-score are 95.7921\%, 99.0317\%, 99.4988\%, and 92.0855\%, respectively for the full dataset. In these two tables, the results for anomaly are higher than normal because there are more data for anomaly than normal and the ANN could fit better to the majority class. In addition, the accuracy is 99.7942\% for the subset while 99.9326\% for the full dataset. This is expected because with more data used for training, ANN has better performance in terms of precision, recall, and f1-score. This is because there are many parameters (weights between neurons) in ANN, it can better fit the problem with more data used.

\begin{table*}[htbp]
\begin{center}
\caption{Binary classification based on the subset dataset.}
\begin{tabular}{ccccc}
\\ \hline
\multicolumn{5}{c}{Source : Subset Dataset}\\
\multicolumn{5}{c}{Accuracy : 0.997942}\\ \hline
& Precision & Recall & f1-score & Support\\ \hline
Macro & 0.948759 & 0.900779 & 0.923214 & 2499\\
Weighted & 0.983063 & 0.983593 & 0.98313 & 2499\\
Anomaly & 0.987743 & 0.994891 & 0.991304 & 2349\\ 
Normal & 0.909774 & 0.806667 & 0.855124 & 150\\
\hline
\end{tabular}
  \label{tab:binarySubDataset}
  \end{center}
\end{table*}

\begin{table*}[htbp]
\begin{center}
\caption{Binary classification based on full dataset.}
\begin{tabular}{ccccc}
\\ \hline
\multicolumn{5}{c}{Source : Full Dataset}\\
\multicolumn{5}{c}{Accuracy : 0.999326}\\ \hline
& Precision & Recall & f1-score & Support\\ \hline
Macro & 0.984392 & 0.934559 & 0.957921 & 156354\\
Weighted & 0.990477 & 0.990573 & 0.990317 & 156354\\
Anomaly & 0.991354 & 0.998648 & 0.994988 & 146503\\
Normal & 0.977431 & 0.87047 & 0.920855 & 9851\\
\hline
\end{tabular}
  \label{tab:binaryFullDataset}%
  \end{center}
\end{table*}

\subsection {Category Level Attack Classification}
Tables \ref{tab:categorySubDataset} and \ref{tab:categoryFullDataset} show the category classification based on the subset dataset and full dataset, respectively. In Table \ref{tab:categorySubDataset}, the f1-scores for macro f1-score, weighted f1-score, and the f1-scores for DoS, MITM, ARP, Spoofing, Mirai, Normal and Scan  are 78.9534\%, 88.8707\%, 99.5851\%, 34.1463\%, 94.0527\%, 90\%, and 76.9831\%, respectively. In Table \ref{tab:categoryFullDataset}, the macro f1-score, weighted f1-score, and the f1-scores for DoS, MITM, ARP, Spoofing, Mirai, Normal and Scan  are 92.2369\%, 94.7741\%, 99.8712\%, 86.7487\%, 96.6369\%, 92.4553\%, and 85.4726\%, respectively. The accuracy in Table \ref{tab:categorySubDataset} is 98.8722\% for the subset while 99.7283\% in Table \ref{tab:categoryFullDataset} for the full dataset. This is because ANN could extract more hidden knowledge with more data used for training.

\begin{table*}[htbp]
\begin{center}
\caption{Category classification based on subset dataset.}

\begin{tabular}{ccccc}
\\ \hline
\multicolumn{5}{c}{Source : Subset Dataset}\\
\multicolumn{5}{c}{Accuracy : 0.988722}\\ \hline
& Precision & Recall & f1-score & Support\\ \hline
Macro & 0.819231 & 0.792651 & 0.789534 & 2499\\
Weighted & 0.897955 & 0.892757 & 0.888707 & 2499\\
DoS & 0.991736 & 1 & 0.995851 & 240\\
MITM ARP Spoofing & 0.530303 & 0.251799 & 0.341463 & 139\\
Mirai & 0.955763 & 0.925769 & 0.940527 & 1657\\
Normal & 0.969231 & 0.84 & 0.9 & 150\\
Scan & 0.649123 & 0.945687 & 0.769831 & 313\\
\hline
\end{tabular}
  \label{tab:categorySubDataset}%
  \end{center}
\end{table*}

\begin{table*}[htbp]
\begin{center}
\caption{Category classification based on full dataset.}
\begin{tabular}{ccccc}
\\ \hline
\multicolumn{5}{c}{Source : Full Dataset}\\
\multicolumn{5}{c}{Accuracy : 0.997283}\\ \hline
& Precision & Recall & f1-score & Support\\ \hline
Macro & 0.931344 & 0.915716 & 0.922369 & 156354\\
Weighted & 0.949057 & 0.947363 & 0.947741 & 156354\\
DoS & 0.999389 & 0.998035 & 0.998712 & 14760\\
MITM ARP Spoofing & 0.887314 & 0.848526 & 0.867487 & 8853\\
Mirai & 0.967635 & 0.965106 & 0.966369 & 104057\\
Normal & 0.985411 & 0.870775 & 0.924553 & 9851\\
Scan & 0.816972 & 0.89614 & 0.854726 & 18833\\
\hline
\end{tabular}
  \label{tab:categoryFullDataset}%
  \end{center}
\end{table*}

\subsection{Subcategory Level Attack Classification}
Tables \ref{tab:subCatSubDataset} and \ref{tab:subCatFullDataset} show the Subcategory classification based on the subset dataset and full dataset, respectively. As we can see from the tables, the accuracy in Table \ref{tab:subCatSubDataset} is 95.4654\% while 97.767\% in Table \ref{tab:subCatFullDataset}. In Table \ref{tab:subCatSubDataset}, f1-scores for macro, weighted, and all nine subcategory attacks are 55.5961\%, 62.6079\%, 100\%, 39.6476\%, 22.4599\%, 27.6029\%, 68.3642\%, 77.9501\%, 87.2852\%, 12.6761\% and 55.3792\%, respectively. In Table \ref{tab:subCatFullDataset}, f1-scores for macro f1-score, weighted f1-score, and labels 1 to 9 f1-scores are 63.6242\%, 69.2957\%, 99.8508\%, 87.6577\%, 29.7254\%, 9.1764\%, 83.7163\%, 77.6855\%, 91.6888\%, 11.7081\%, and 72.3184\% , respectively. The results for DoS - Synflooding are higher than MITM ARP Spoofing because there are more data for DoS - Synflooding  than DoS - Synflooding and the ANN could fit better to the majority class. The results in Table \ref{tab:subCatFullDataset} are better than that in Table \ref{tab:subCatSubDataset} is because there are many parameters (weights between neurons) in ANN, it can better fit the problem with more data used.

In summary, with more data used for training, ANN achieves better results based on the full dataset in Tables \ref{tab:binaryFullDataset}-\ref{tab:categoryFullDataset} reinforces the optimal hyperparameter selection from the subset in Tables \ref{tab:binarySubDataset}-\ref{tab:subCatSubDataset}.

\begin{table*}[htbp]
\begin{center}
\caption{Subcategory classification based on subset dataset.}
\begin{tabular}{ccccc}
\\ \hline
\multicolumn{5}{c}{Source : Subset Dataset}\\
\multicolumn{5}{c}{Accuracy : 0.954654}\\ \hline
& Precision & Recall & f1-score & Support\\ \hline
Macro & 0.570248 & 0.553568 & 0.545961 & 2499\\
Weighted & 0.628303 & 0.641857 & 0.626079 & 2499\\
DoS-Synflooding & 1 & 1 & 1 & 240\\
MITM ARP Spoofing & 0.511364 & 0.323741 & 0.396476 & 139\\
Mirai-Ackflooding & 0.259259 & 0.198113 & 0.224599 & 212\\
Mirai-HTTP Flooding & 0.3 & 0.255605 & 0.276029 & 223\\
Mirai-Hostbruteforceg & 0.680312 & 0.687008 & 0.683643 & 508\\
Mirai-UDP Flooding & 0.751625 & 0.809524 & 0.779501 & 714\\
Normal & 0.900709 & 0.846667 & 0.872852 & 150\\
Scan Hostport & 0.3 & 0.080357 & 0.126761 & 112\\
Scan Port OS & 0.428962 & 0.781095 & 0.553792 & 201\\
\hline
\end{tabular}
  \label{tab:subCatSubDataset}%
  \end{center}
\end{table*}

\begin{table*}[htbp]
\begin{center}
\caption{Subcategory classification based on full dataset.}
\begin{tabular}{ccccc}
\\ \hline
\multicolumn{5}{c}{Source : Full Dataset}\\
\multicolumn{5}{c}{Accuracy : 0.97767}\\ \hline
& Precision & Recall & f1-score & Support\\ \hline
Macro & 0.700202 & 0.637441 & 0.626142 & 156354\\
Weighted & 0.703244 & 0.731897 & 0.691957 & 156354\\
DoS-Synflooding & 0.999457 & 0.997561 & 0.998508 & 14760\\
MITM ARP Spoofing & 0.890365 & 0.86321 & 0.876577 & 8853\\
Mirai-Ackflooding & 0.321678 & 0.276278 & 0.297254 & 13852\\
Mirai-HTTP Flooding & 0.416138 & 0.051568 & 0.091764 & 14001\\
Mirai-Hostbruteforceg & 0.792068 & 0.887703 & 0.837163 & 30553\\
Mirai-UDP Flooding & 0.692747 & 0.884208 & 0.776855 & 45651\\
Normal & 0.960739 & 0.876865 & 0.916888 & 9851\\
Scan Hostport & 0.590615 & 0.064981 & 0.117081 & 5617\\
Scan Port OS & 0.638015 & 0.834594 & 0.723184 & 13216\\
\hline
\end{tabular}
  \label{tab:subCatFullDataset}%
  \end{center}
\end{table*}

\begin{figure}
\begin{subfigure}{.5\textwidth}
  \centering
  \includegraphics[width=1\linewidth]{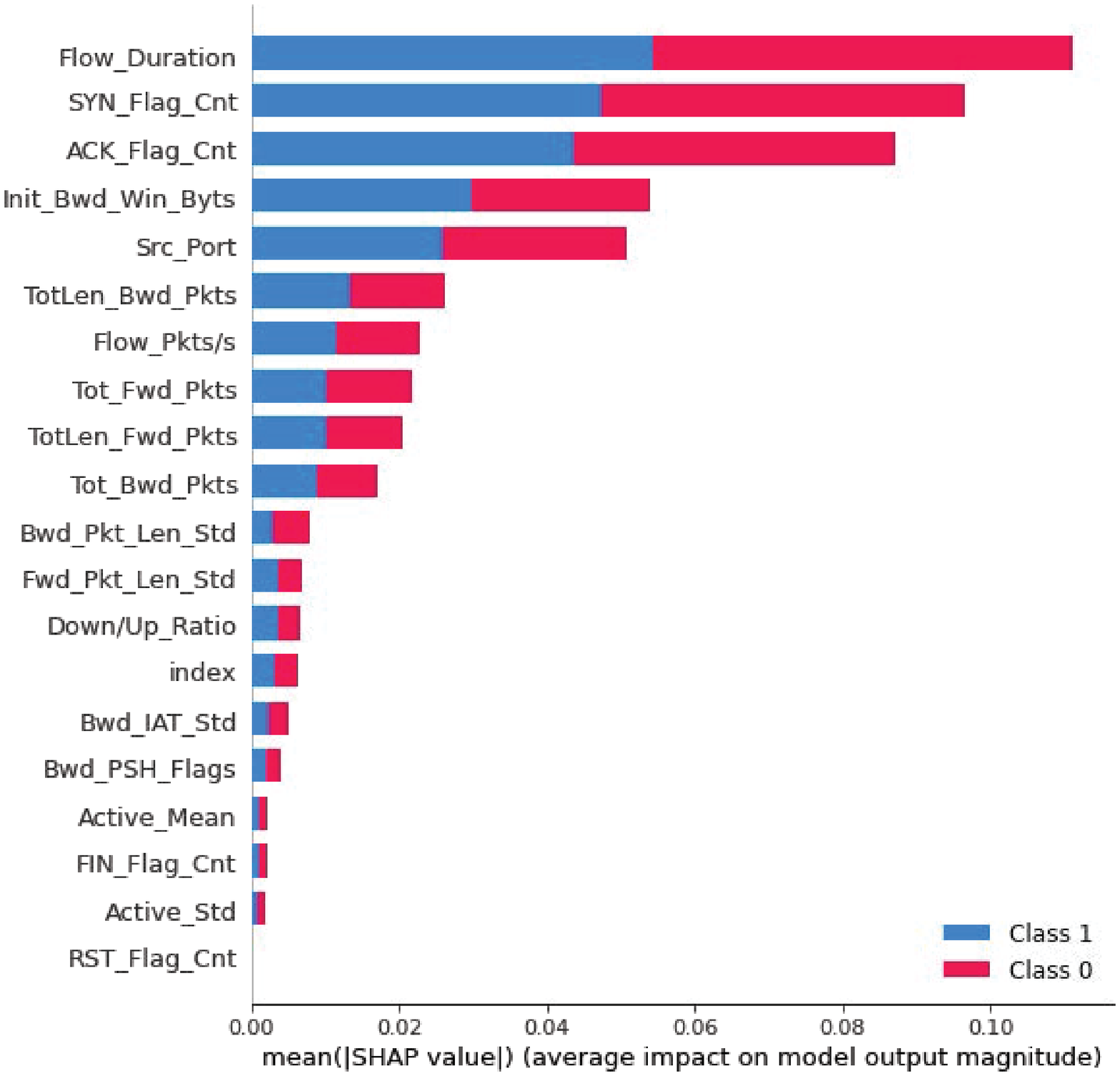}
  \caption{Feature importance plot.}
\end{subfigure}%
\begin{subfigure}{.5\textwidth}
  \centering
  \includegraphics[width=1\linewidth]{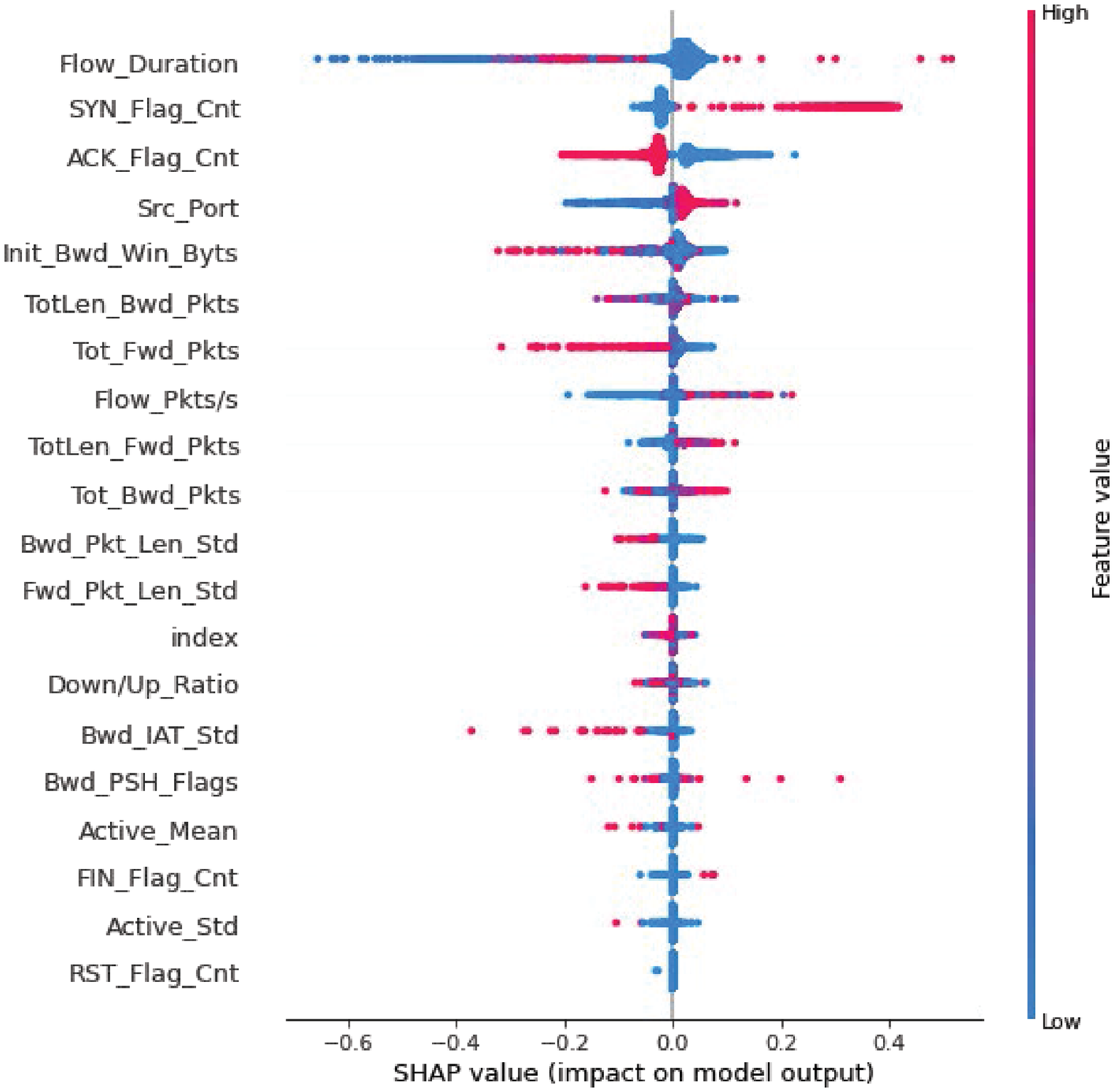}
  \caption{Summary plot.}
\end{subfigure}
\caption{Shap plots for binary classification based on the subset dataset.}
\label{fig:shap:binary}
\end{figure}
\subsection{Comparison of Our System with Other Related Work}
We have compared performance of our approach with recent existing work using the same data and the summary is presented in Table \ref{tab:comparison}. It is evident from the table that only Ullah and Machmoud \cite{Ullah2020} has considered subcategory level of attack classification for this dataset and the accuracy and F score is 88\% with decision tree method. Most of the work that considered IoTID20 dataset did not classify attacks at subcategory level. As discussed in the Results section (Section \ref{results}) we have performed comprehensive set of experiments and achieved high accuracy for all three levels of attack classification.

\begin{table*}[htbp]
 \centering
\caption{Comparison With Other Related Work}
\label{tab:comparison}
  \begin{tabular}{m{40pt}<{\centering} m{80pt}<{\centering} m{60pt}<{\centering} m{90pt}<{\centering}}
 \hline
Research work &  Machine Learning Approaches used & Classification label & Best Result \\ \hline
Ullah and Mahmound \cite{Ullah2020} & Multiple & Binary, Category and Subcategory & Subcategory = 88\% F-Score, 88\% Accuracy (DT) \\ \hline
Qaddoura et al. \cite{qaddoura2021} & Single Hidden Layer Feed-Forward Neural Network (SLFN) & Binary & 98.42\% Accuracy \\ \hline
Farah \cite{farah2020} & Multiple & Binary and Category & Binary = 0.96 AUC, Category = 0.98 AUC (Ensemble)  \\ \hline
Krishnan, Nayaz and Liu \cite{krishnannetwork2021} & SVC, XGBoost and RF & Binary & 99.97 Accuracy (XGBoost) \\ \hline
MAMID (proposed approach) & ANN with optimal Hyperparameters & Binary, Category and Subcategory & 99.9\%, 99.7\%, and 97.7\% Accuracy \\ \hline
\end{tabular}
\end{table*}

\subsection{Explainability Analysis}

As discussed previously, explainability is an important factor when it comes to adaptation of complex machine learning models by end users. MAMID is designed for smart home scenario where users trust in the system is crucial for users to adapt this security system in their homes. In this section we briefly discuss the explainability of the models based on the subset dataset.

The SHAP plots shown in Figure \ref{fig:shap:binary} show the feature importance and summary plot for binary classification  based on subset dataset. In both the plots, the features are placed on the Y-Axis according to their importance. 

The feature importance plot shows how each feature impacts the detection of the classes, in this case for binary classification Class 1 is normal (blue) and Class 2 is attack (red). The impact of every feature on the detection of the attacks show that Flow\_Duration has the highest impact on detection of these classes. 

\begin{figure}
\begin{subfigure}{.5\textwidth}
  \centering
  \includegraphics[width=1\linewidth]{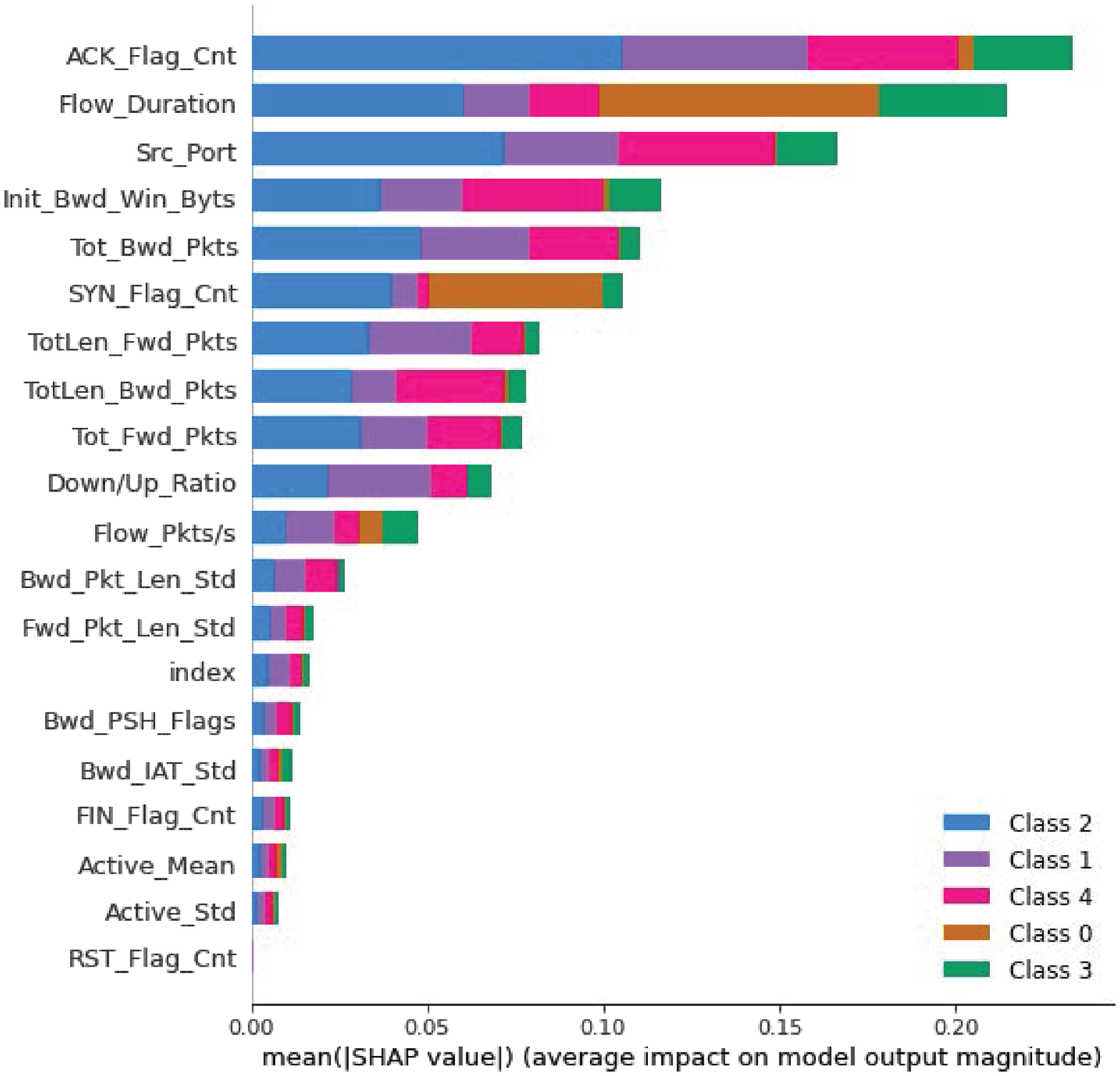}
  \caption{Feature importance plot.}
\end{subfigure}%
\begin{subfigure}{.5\textwidth}
  \centering
  \includegraphics[width=1\linewidth]{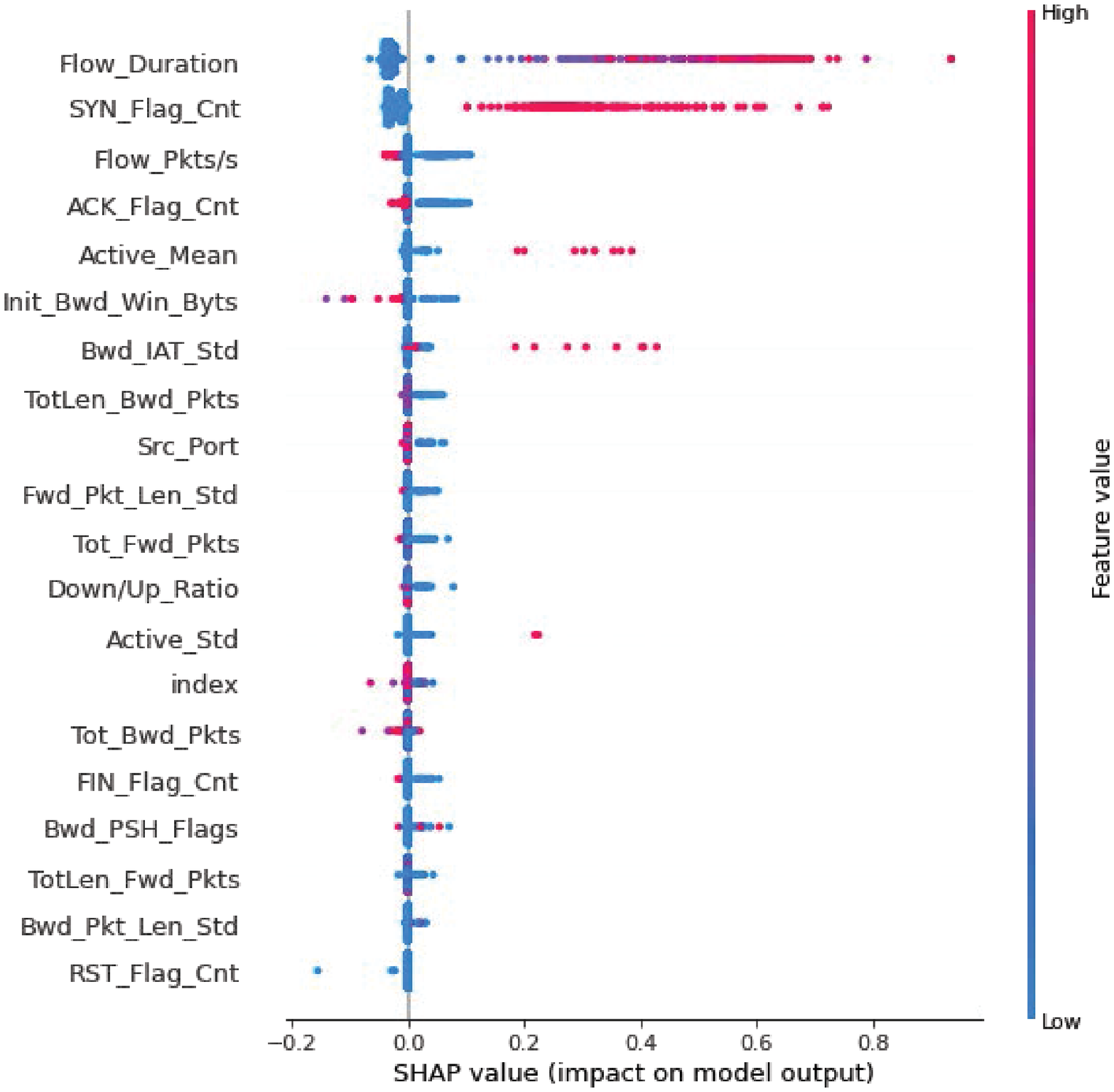}
  \caption{Summary plot.}
\end{subfigure}
\caption{Shap plots for category classification based on the subset dataset.}
\label{fig:shap:category}
\end{figure}

The summary plot shows the positive and negative relationships of the predictors with the target variable. The value on the X-Axis in summary plot shows whether the effect of that value is associated with a higher or lower prediction, the color shows high (in red) or low (in blue) for that observation. The low level of the Flow\_Duration has a low and negative impact on the attack detection. Similarly, the feature importance plot and summary plot for category classification are shown in Figure\ref{fig:shap:category}. 

\begin{figure}
  \centering
  \includegraphics[width=1\linewidth]{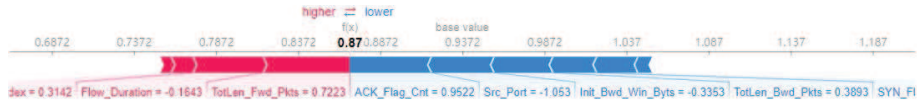}
  \caption{Force plot for binary classification based on the subset dataset.}
\label{fig:force-binary}
\end{figure}

The force plot shown in Figure \ref{fig:force-binary} shows the effect of different features on a particular predicted value for binary classification based on subset dataset. The given plot shows the baseline and the average predicted probability for a particular value to be classified as an attack is 0.07. The data sample has low predicted value as the increasing effect of TotLen\_Fwd\_pkts field is offset by decreasing effect of ACK\_Flag\_Cnt field.    

\section{Conclusion}
\label{conclusion}
Accurate detection of intrusion is very important in cybersecurity. In this study, we used ANN for intrusion detection based on hyperparameter optimisation. To examine the effectiveness of different hyperparameters for ANN, experiments were conducted, including changing the number of epochs, batch size, number of neurons, optimiser, activation for input and hidden layers, and activation for the output layer. To validate our proposed model, a dataset with three different cases was considered: (1) Label classification (binary), (2) Category classification (5 classes), and (3) Subcategory classification (9 classes). The experimental results revealed that our proposed model with hyperparameter optimisation has high performance for intrusion detection with 99.9\%, 99.7\%, and 97.7\% accuracy for Label, Category, and Subcategory classification. Although we only tested our proposed model in intrusion detection, it can also be applied to other cybersecurity problems (e.g. malicious website domain detection) and also other areas (e.g. disease detection and anomaly detection). In the future work, we will investigate more about the ANN structure, including automated selection of the right hyperparameters to prevent over-fitting problems. We will also explore different feature selection or feature extraction options for our proposed model.

\bibliography{reference}

\begin{thebibliography}{10}

\bibitem{lee2020}
I~Lee.
\newblock {\em Securing the Internet of Things: Concepts, Methodologies, Tools,
  and Applications}.
\newblock IGI Global, USA, 2020.

\bibitem{yang2017survey}
Yuchen Yang, Longfei Wu, Guisheng Yin, Lijie Li, and Hongbin Zhao.
\newblock A survey on security and privacy issues in internet-of-things.
\newblock {\em IEEE Internet of Things Journal}, 4(5):1250--1258, 2017.

\bibitem{thamilarasu2019deepLearning}
Geethapriya Thamilarasu and Shiven Chawla.
\newblock Towards deep-learning-driven intrusion detection for the internet of
  things.
\newblock {\em Sensors}, 19(9), 2019.

\bibitem{dolan2020Proactively}
Andy Dolan, Indrakshi Ray, and Suryadipta Majumdar.
\newblock Proactively extracting iot device capabilities: An application to
  smart homes.
\newblock In Anoop Singhal and Jaideep Vaidya, editors, {\em Data and
  Applications Security and Privacy XXXIV}, pages 42--63, Cham, 2020. Springer
  International Publishing.

\bibitem{zhang2000neuralnetwork}
G.~Zhang.
\newblock Neural networks for classification: a survey.
\newblock {\em IEEE Trans. Syst. Man Cybern. Part C}, 30:451--462, 2000.

\bibitem{dreiseitl2002LogisticRA}
S.~Dreiseitl and L.~Ohno-Machado.
\newblock Logistic regression and artificial neural network classification
  models: a methodology review.
\newblock {\em Journal of biomedical informatics}, 35 5-6:352--9, 2002.

\bibitem{choras20217ID}
Michał Choraś and Marek Pawlicki.
\newblock Intrusion detection approach based on optimised artificial neural
  network.
\newblock {\em Neurocomputing}, 452:705--715, 2021.

\bibitem{Ullah2020}
I.~Ullah and Q.~Mahmoud.
\newblock A scheme for generating a dataset for anomalous activity detection in
  iot networks.
\newblock In {\em Canadian Conference on AI}, 2020.

\bibitem{islam2020deep}
Fabliha~Bushra Islam, Rubina Akter, Dong-Seong Kim, and Jae-Min Lee.
\newblock Deep learning based network intrusion detection for industrial
  internet of things.
\newblock pages 418--421, 2020.

\bibitem{tsaih2018ann}
Rua-Huan Tsaih, Shin-Ying Huang, Mao-Ci Lian, and Yennun Huang.
\newblock Ann mechanism for network traffic anomaly detection in the concept
  drifting environment.
\newblock In {\em 2018 IEEE Conference on Dependable and Secure Computing
  (DSC)}, pages 1--6, 2018.

\bibitem{taher2019NID}
Kazi~Abu Taher, Billal Mohammed Yasin~Jisan, and Md.~Mahbubur Rahman.
\newblock Network intrusion detection using supervised machine learning
  technique with feature selection.
\newblock In {\em 2019 International Conference on Robotics,Electrical and
  Signal Processing Techniques (ICREST)}, pages 643--646, 2019.

\bibitem{hodo2016ann}
Elike Hodo, Xavier Bellekens, Andrew Hamilton, Pierre-Louis Dubouilh, Ephraim
  Iorkyase, Christos Tachtatzis, and Robert Atkinson.
\newblock Threat analysis of iot networks using artificial neural network
  intrusion detection system.
\newblock In {\em 2016 International Symposium on Networks, Computers and
  Communications (ISNCC)}, pages 1--6, 2016.

\bibitem{subba2016ann}
Basant Subba, Santosh Biswas, and Sushanta Karmakar.
\newblock A neural network based system for intrusion detection and attack
  classification.
\newblock In {\em 2016 Twenty Second National Conference on Communication
  (NCC)}, pages 1--6, 2016.

\bibitem{unsw15}
Nour Moustafa and Jill Slay.
\newblock Unsw-nb15: a comprehensive data set for network intrusion detection
  systems (unsw-nb15 network data set).
\newblock In {\em 2015 Military Communications and Information Systems
  Conference (MilCIS)}, pages 1--6, 2015.

\bibitem{botiot}
Nickolaos Koroniotis, Nour Moustafa, Elena Sitnikova, and Benjamin Turnbull.
\newblock Towards the development of realistic botnet dataset in the internet
  of things for network forensic analytics: Bot-iot dataset.
\newblock {\em Future Generation Computer Systems}, 100:779--796, 2019.

\bibitem{Maniriho2020}
Pascal Maniriho, Ephrem Niyigaba, Zephanie Bizimana, Valens Twiringiyimana,
  Leki~Jovial Mahoro, and Tohari Ahmad.
\newblock Anomaly-based intrusion detection approach for iot networks using
  machine learning.
\newblock In {\em 2020 International Conference on Computer Engineering,
  Network, and Intelligent Multimedia (CENIM)}, pages 303--308, 2020.

\bibitem{qaddoura2021}
Raneem Qaddoura, Ala Al-Zoubi, Iman Almomani, and Hossam Faris.
\newblock A multi-stage classification approach for iot intrusion detection
  based on clustering with oversampling.
\newblock {\em Applied Sciences}, 11(7), 2021.

\bibitem{farah2020}
Anjum Farah.
\newblock {\em Cross Dataset Evaluation for IoT Network Intrusion Detection}.
\newblock PhD thesis, 2020.

\bibitem{mehlig2019artificial}
Bernhard Mehlig.
\newblock Artificial neural networks.
\newblock {\em arXiv e-prints}, pages arXiv--1901, 2019.

\bibitem{optimisers}
Sanket Doshi.
\newblock Various optimization algorithms for training neural network, 2019.

\bibitem{afaq2020S}
Saahil Afaq and Smitha Rao.
\newblock Significance of epochs on training a neural network.
\newblock {\em International Journal of Scientific \& Technology Research},
  9:485--488, 2020.

\bibitem{samuel2017}
Samuel~L. Smith, Pieter{-}Jan Kindermans, and Quoc~V. Le.
\newblock Don't decay the learning rate, increase the batch size.
\newblock {\em CoRR}, abs/1711.00489, 2017.

\bibitem{hayou2019}
Soufiane Hayou, Arnaud Doucet, and Judith Rousseau.
\newblock On the impact of the activation function on deep neural networks
  training, 2019.

\bibitem{Lundberg2017}
Scott Lundberg and Su-In Lee.
\newblock A unified approach to interpreting model predictions.
\newblock pages 1--10, 12 2017.

\bibitem{Huang2020}
Sheng-Wen Huang, Huey-Pin Tsai, Su-Jhen Hung, Wen-Chien Ko, and Jen-Ren Wang.
\newblock Assessing the risk of dengue severity using demographic information
  and laboratory test results with machine learning.
\newblock {\em PLOS Neglected Tropical Diseases}, 14:e0008960, 12 2020.

\bibitem{krishnannetwork2021}
Sundar Krishnan, Ashar Neyaz, and Qingzhong Liu.
\newblock Iot network attack detection using supervised machine learning.
\newblock {\em International Journal of Artificial Intelligence and Expert
  Systems (IJAE)}, 10(2):18--32, June 2021.

\end{thebibliography}
\bibliographystyle{unsrt}

\end{document}